\documentclass[12pt]{iopart}

\usepackage{iopams} 
\usepackage{cite}
\usepackage{graphicx}
\usepackage{amssymb}
\usepackage[geometry]{ifsym}
\usepackage{caption}
\usepackage{subcaption}
\usepackage[dvipsnames]{xcolor}
\begin{document}

\title[Jamming of active particles in narrow pores]{Jamming of active particles in narrow pores: Implications for ratchet effect and diffusion coefficient}

\author{Šimon Pajger$^{1,2}$ and František Slanina$^{1}$}

\address{$^{1}$Institute of Physics, Czech Academy of Sciences, Na Slovance 2, CZ-18200 Praha, Czech Republic}
\address{$^2$Faculty of Mathematics and Physics, Charles University, Ke Karlovu 2027/3, CZ-12116 Praha, Czech Republic}
\ead{pajger@fzu.cz}
\vspace{10pt}
\begin{indented}
\item[]2025
\end{indented}

\begin{abstract}
We study the behavior of colloidal active particles interacting via steric repulsion in various quasi-1D geometries. We mainly focus on active particles with high Péclet number. We discuss 3 phenomena closely tied to those systems: motility-induced phase separation (or dynamical freezing), ratchet effect (which takes place if the geometry has broken spatial symmetry), and the enhanced diffusion coefficient. We study those particles using numerical simulations employing an ASEP-like model. Besides direct numerical simulations we study the model by mean-field approximation and by solving the coarse-grained hydrodynamic equations.
\end{abstract}

%
%
%
%
%

\section{Introduction}

Active particles, both occurring in the natural world
\cite{dom_cis_cha_gol_kes_04,mak_rat_sym_jag_lee_ner_06,sok_ara_09}
and prepared  artificially
\cite{paxton_etal_04,how_jon_rya_gou_vaf_gol_07,wal_mul_13,pal_sac_ste_pin_cha_13,nar_ram_men_07}, 
are a subject of intense research in the last couple of decades
\cite{bas_mar_09,ramaswamy_10,rom_bar_ebe_lin_12,mar_joa_ram_liv_pro_rao_sim_13,bec_dil_low_rei_vol_vol_16,fod_mar_18}. The possible
applications range from material science to drug delivery (see e.g.
\cite{gho_xu_gup_gra_20,wan_zho_21}). These potential applications, in
turn, create a need for a theoretical description of the behavior of
active colloidal particles.

Active particles can convert energy from their
surroundings into directed motion, driving themselves
out of equilibrium. Collections of interacting active particles form
genuinely non-equilibrium systems,  often called `active matter'. In these
systems, energy is dissipated at the scale of individual constituents,
as each active particle has irreversible dynamics.
The direct consequence of the active propulsion is the enhancement of the effective
diffusion coefficient, as demonstrated already in the early
experiments, see e.g.  \cite{how_jon_rya_gou_vaf_gol_07}.
Compared to passive particles,
active particles hold their direction of motion more consistently,
because while the autocorrelation of velocities in the case of passive
particles is caused mainly by inertia (which is in the overdamped
dynamics negligible), active particles are also driven by their
internal mechanism. The direction of this driving decorrelates
generally on a much larger timescale. This causes active particles to
have a higher diffusion coefficient than their passive counterparts by
several orders of magnitude. In interacting systems, more detailed information can be obtained
from tracer diffusion, using either active or passive particles as
tracers \cite{ban_jac_cat_22}.

However, the most important difference of active particles compared 
to their passive counterparts consists in the 
emergence of phenomena which are absent in systems which follow the
principles of equilibrium statistical physics. The most striking phenomenon of this kind is the motility-induced phase separation
(MIPS) or dynamical freezing
\cite{tai_cat_08,cat_tai_15,byr_sol_tai_zha_23},
which resembles the usual phase
coexistence of low-density and high-density phases in equilibrium
systems. However, contrary to the latter, it occurs also when
particles interact purely repulsively. Moreover, the
common-tangent construction determining the density of coexisting
phases in equilibrium is replaced by much more complex conditions
\cite{byr_sol_tai_zha_23} and
in many cases can be obtained only by direct numerical
computation. Therefore, the physics behind MIPS is highly
model-specific and to a large extent open to investigations.

While MIPS occurs in unbounded space, presence of geometric
constraints in the form of boundary conditions  brings about further
phenomena which are absent in equilibrium. Notably, the active
particles accumulate at walls and moreover the accumulated density
depends on local curvature of the wall
\cite{pot_sta_12,fil_bas_hag_14} and can lead to emergence of
macroscopic particle
current due to the ratchet effect. The necessary condition is breaking
the mirror symmetry of the confining geometry. In passive systems, it
appears due to periodic external driving \cite{reimann_02}, but in
active systems it appears spontaneously. Practically, this  feature
was observed experimentally in rectification structures either in 
quasi-one-dimensional
\cite{gal_key_cha_aus_07,mah_cam_bis_kom_cha_soh_hud_09,sok_apo_grz_ara_10,dil_ang_del_ruo_10}  
or two-dimensional
\cite{vol_but_vog_kum_bec_11,kai_pes_sok_etal_14,gui_jey_berd_etal_14} 
geometries. Theoretical investigations followed using both simulations
and analytical approaches. Without any ambition of representativeness,
we can mention
e.g. Refs. \cite{wan_ols_nus_rei_08,ang_dil_ruo_09,ber_etal_13,gho_mis_mar_nor_13,ao_gho_li_sch_han_mar_14,ai_che_he_li_zho_13,kou_mag_dil_14,ai_16}
or the
review \cite{ols_rei_17}.

The ratchet effect, as a single-particle phenomenon, will be of course
altered when particle interactions are taken into account. With active
particles, the most important change comes from the emergence of MIPS,
which can lead to jamming that hinders or fully inhibits the expected
ratchet current. The main question we shall address in this work is
how the MIPS affects the ratchet current in simple quasi-one-dimensional
geometries.

To this end, we apply a generalized asymmetric simple exclusion process (ASEP) model
\cite{ari_kra_mal_14,hum_kot_net_sla_20}, also called partial
exclusion process in
different contexts \cite{sch_san_94,tai_kur_lec_08,tho_tai_cat_bly_10}.
Here, the overdamped continuous dynamics of active Brownian
particles is mapped onto discrete dynamics of an active lattice gas
using a heuristic argument that we call ``local mixing approximation''
\cite{sla_kot_net_22,sla_kot_25}. It consists in discretizing the space into
disjoint cells and emulating the steric repulsion
of active particles by exclusion constraint requiring that each cell
can accommodate at most $k$ particles. Therefore, we follow the line
of active lattice gases, explored earlier
\cite{tho_tai_cat_bly_10,sla_kot_net_22,sla_kot_25,per_kla_deu_vos_11,sol_tai_13,sot_gol_14,kou_eri_bod_tai_18,dit_spe_vir_21,agr_ro_kaf_lec_23,mas_eri_jac_bru_23,ros_dia_dic_24}.

In a recent article \cite{sla_kot_25} we studied similar model in
two-dimensional setting. Here we simplify the geometry to
quasi-one-dimensional one, which enables us to go deeper into
details and also makes easier the comparison to continuous limit
developed by us recently \cite{sla_kot_23}.
The paper is organized as follows. In the next section we
define the model and in Section \ref{sec:D_jam} we explore the influence of MIPS on
diffusion of a tracer particle. In Section \ref{sec:CGrho} we show the phase diagram
with P\'eclet number as governing quantity. In Sections \ref{S:JT}, \ref{S:c-d} and \ref{sec:hydro} we
investigate the metastable and stationary currents and compare the
results to continuous approximations. The section \ref{sec:conclusion} draws conclusions
from the results.

\section{Description of the model}\label{model}

We use a quasi-1D channel with periodic boundary conditions and model interaction between particles just as steric repulsion. To do that we discretize our channel into several cells and for each cell, we use the local mixing approximation \cite{sla_kot_net_22,sla_kot_25}. It consists of the assumption, that the number of particles in the given cell is sufficient to describe the state of that cell itself and the precise position of the particle within the cell is immaterial. Then the steric repulsion between the particles is reduced to a condition, that in each cell there might be at most $k$ particles at the time.

We will model the evolution of the system as a continuous-time Markov process. We thus need rates for the different possible steps that can be made. Each active particle has its own direction of propulsion. We denote the rate of jumps in the direction of propulsion as $a$ and the rate of jumps in the opposite direction as $b$. These two rates combine contributions from both active propulsion and passive diffusion. Namely, the symmetric part $(a+b)/2$ quantifies the symmetric (passive) diffusion, while the modulus of the asymmetric part $|a-b|$ quantifies the propulsion velocity, with direction set by the internal degree of freedom of the particle. We denote the rate of switching the direction $\alpha$. In our model we define the Péclet number Pe $= (a-b)/\sqrt{\alpha (a+b)}$. The Péclet number quantifies the relative importance of active particle propulsion, our work mostly focuses on the regime when the Pe is high.

\begin{figure}[ht]\centering
\includegraphics[width=100mm]{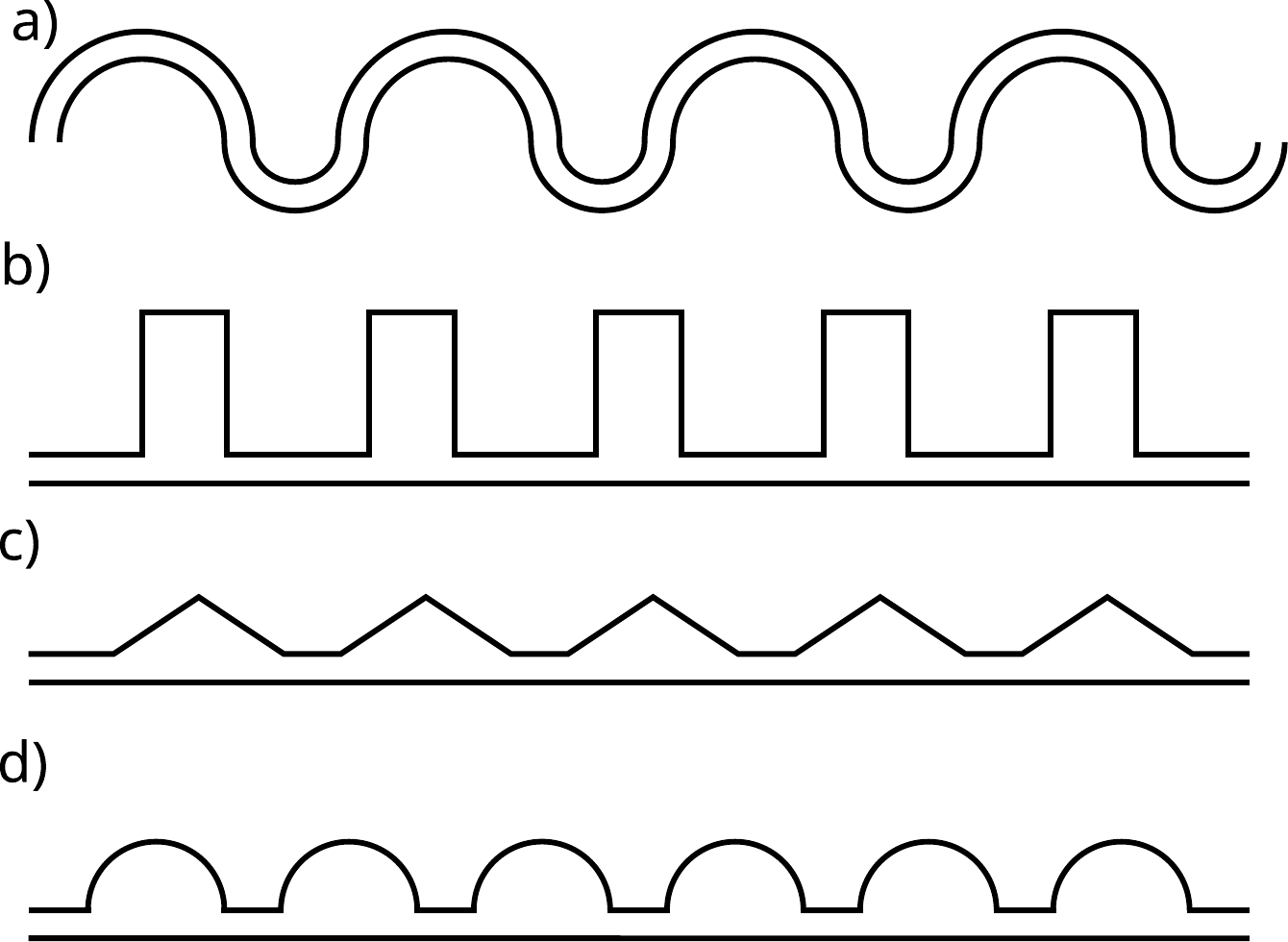}
\caption{Sketch of the geometries illustrating the shapes similar to
    those used in several experiments: a) for Ref. \cite{dic_iri_tom_ton_07}
    b) for Ref. \cite{lee_cho_par_11}, c) and d) for Ref. \cite{zha_li_li_ali_13}.}
\label{fig:schemes}
\end{figure}

Let us now turn to the geometry of the channel. The experimental motivation for our work is based on microfluidic
devices, where colloidal particles flow in channels of various complex
shapes. The typical length scale of the particles is $1\,\mu\mathrm{m}$
and the aperture of the channels is appropriately larger. For a review
on Microfluidics, see e.g. \cite{squ_qua_05}. The shapes of the
channels are designed to facilitate particle sorting and
directing. Without any ambition of representativeness, we can cite
e.g. experimental works
\cite{dic_iri_tom_ton_07,lee_cho_par_11,zha_li_li_ali_13}. We sketch
the geometries that are similar to those used in these works in
Fig. \ref{fig:schemes}. The 
common feature of these shapes is the appearance of periodically
placed deviations, which may act as temporal traps for the movement of
the particles. The geometries used by us are theoretical idealizations
of such shapes of microfluidic devices.

\begin{figure}[ht]\centering
\includegraphics[width=140mm]{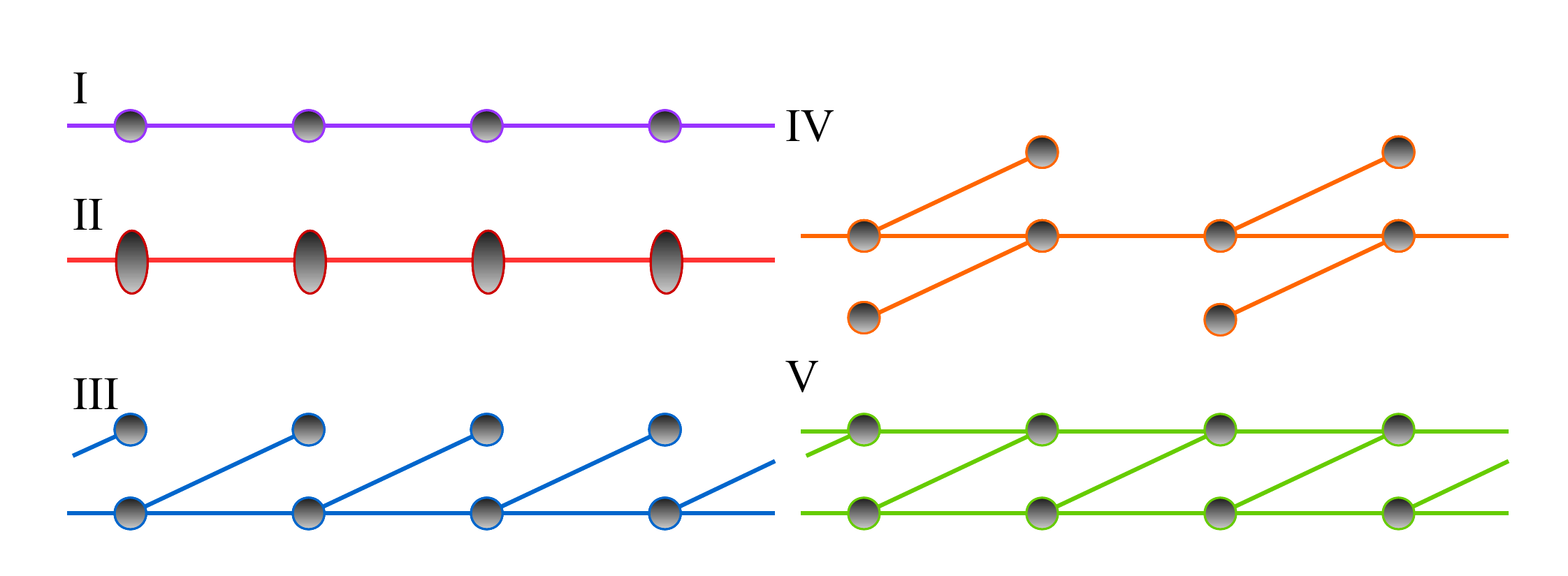}
\caption{Sketches of different geometries. Geometries I and II are smooth tubes with $L$ cells and the capacity of one cell $k=2$ and $k=4$, respectively. Geometry III has $2L$ cells, half of which are pockets, all of which are appended to the backbone in the same direction, and the capacity of one cell $k=2$. Geometry IV is analogous to Geometry III, except the directions of pockets are alternating. Geometry V is obtained from Geometry III by connecting pockets into another backbone. All geometries are depicted in this figure at length $L=4$, and we always use periodic boundary conditions.}
\label{fig:geom}
\end{figure}

We shall use several variants of the discretisations of those geometries. The simplest geometry is $L$ cells connected to a single backbone (or to a loop, since we use periodic boundary conditions). We denote it in Fig. \ref{fig:geom} Geometry I and Geometry II. The only difference between them is that Geometry II has the capacity of a single cell $k=4$, while Geometry I has $k=2$. We use both of them, because Geometry I has the same $k$ as geometries III to V, while Geometry II has the same maximal length density as them.

Geometry III is the only geometry we will use that has broken spatial symmetry. In Geometry III, half of the cells are connected to a chain forming a "backbone." But to each of these "backbone cells," there is connected one additional "pocket cell." As the name suggests, the pockets are not connected to any other cell. All pockets are connected to their respective neighbors in the same direction.

Geometry IV is similar to Geometry III, with the only difference being that the directions in which pockets are connected are alternating. Finally, we call Geometry V a "ladder" geometry, which we get from Geometry III by connecting pockets to create another backbone. The sketches of these geometries are in Fig. \ref{fig:geom}.

In the sketches in Fig. \ref{fig:geom} the angles between links connecting cells serve merely the graphical purpose. For particles, there are only 2 directions: forward (rate $a$) and backward (rate $b$), and only two ways in which the particle may be propelled (in Fig. \ref{fig:geom} right or left). However we do need to distinguish between different links going in the same direction for the purposes of the simulation and a verbal description. For this we will utilize the vertical direction, so if a particle does not change its vertical position during its jump, we will say that it stayed in its backbone and if it does change its vertical position, we say it changed ist backbone. Notice, that in a given direction (forward or backward), there are from $0$ to $2$ links exiting each given cell, at most one that does not change the backbone and at most one that changes it, and a particle has the same rate for jumping along either of them.

Let us now describe the evolution of the system. We evolve the system through random sequential updates. We denote the mean waiting time between two independent jump attempts $\tau=(n (2a+2b+\alpha))^{-1}$, where $n$ is the number of particles. In each step, the time increment is from an exponential distribution with mean $\tau$. Then in a given step, we choose a random particle (from a uniform distribution) and then we choose what kind of step this particle will try, by generating a random variable $r$ with a uniform distribution from the interval $(0,2 a + 2b+\alpha)$. If $r \in (0,a)$, the particle tries to jump in its direction of propulsion along its current backbone, and if $r \in (a,2a)$, the particle tries to jump in its direction of propulsion but also change its backbone. Similarly, for jumps against the direction of propulsion are intervals $(2a,2a+b)$ and $(2a+b,2a+2b)$, corresponding to jump along the current backbone and jump to other backbone respectively. However all of those processes mentioned above happen only if they are possible, so if there is a suitable link from the cell chosen particle is in and the cell to which it leads is not full, i.e. contains less than k other particles. If there is not a link corresponding to an attempted step, or such a link leads to an already full cell, the jump does not happen: the particle stays where it was before instead, while the time tics normally. Finally, if $r \in (2a+2b,2a+2b+\alpha)$, the particle reverses its direction of propulsion, this process is always possible. Because of these choices, each jump, that is possible in a current state happens with the correct rate.  

\section{Diffusion coefficient and jamming}\label{sec:D_jam}
We consider the tracer diffusion to calculate the diffusion coefficient and use the following procedure. In a given run and time $t$ we determine for each particle how much it moved from its initial position $x_i(t)$ (including how many wraps around the entire length it had done) and then calculate the variation of these distances over all the particles
\begin{equation}
    \sigma^2(t) = \frac{\sum_{i=1}^n(x_i(t)-\bar x(t))^2}{n} \,,
\end{equation}
where $n$ is the total number of particles in the system. We average the variation $\sigma^2(t)$ over hundreds of independent runs and use the fact, that in limit of $t \rightarrow \infty$, this variation will asymptotically approach $\sigma^2(t)=2 D t$, to calculate the diffusion coefficient $D$.

We typically set $b=0$ because we are mainly interested in the effect of activity, and letting particles jump backward would introduce additional noise, which we are not interested in. We choose $\alpha \ll a$. This choice corresponds to high Péclet number. We initialize the system by putting particles into random cells (that are not yet full) and randomizing their direction of propulsion. Then we let the system to equilibrate by evolving for the time $T_e=10 / \alpha$, and only then do we start to collect data.

\begin{figure}
\includegraphics[width=140mm]{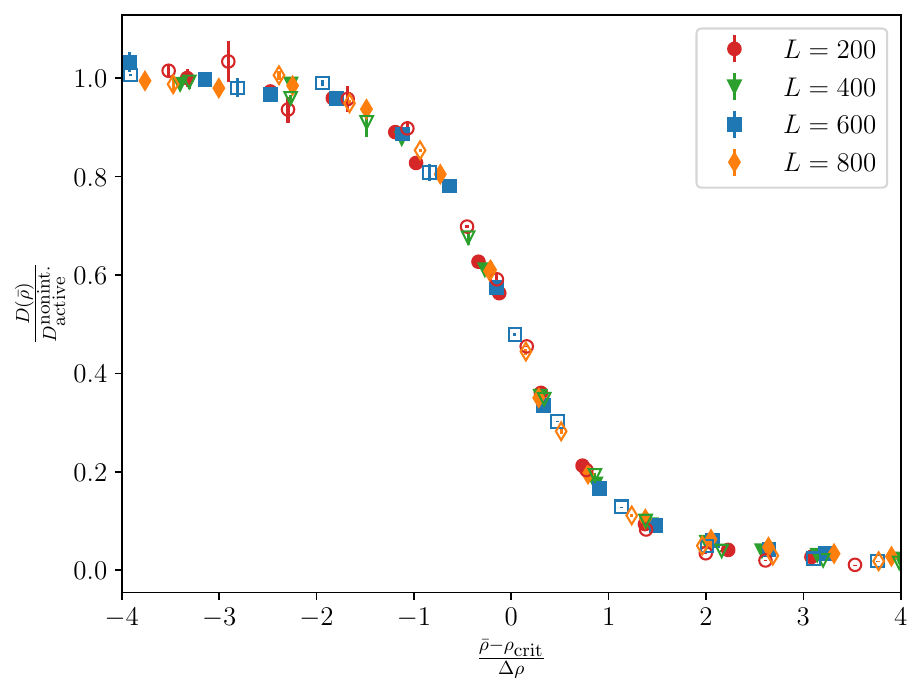}
\caption{Data collapse of diffusion coefficient $D$ as a function of mean density $\bar \rho$. Parameter $\rho_{crit}$ is such a $\bar \rho$ for which $D(\bar \rho)=D_{\rm{active}} ^{\rm{nonint.}}/2$ and parameter $\Delta \rho$ is inversely proportional to derivative of $D$ with respect to $\bar \rho$ at $\rho_{crit}$. We used Geometry I and rates $a=1,b=0$, full symbols are for $\alpha=0.01$, empty ones for $\alpha=0.004$. The sizes of the systems are indicated in the legend.}
\label{fig:D:ex}
\end{figure}

For Geometry I we can see the dependence of $D$ on the average density $\bar \rho$ in Fig. \ref{fig:D:ex}. In order to quantify the contribution to the diffusion coefficient of activity itself, we compare this kind of movement to a single passive particle, which is driven externally in such a way that it undergoes a biased random walk described by rates $a,b$. Such a particle would have the diffusion coefficient $D_{\rm{ passive}}=(a+b)/2$, while noninteracting active particles should have the diffusion coefficient $D_{\rm{active}} ^{\rm{nonint.}}=(a-b)^2/(2 \alpha)+ D_{\rm{passive}}$. For small densities, we indeed get this value. Note that we are measuring the distances in units of cells (which have a similar length scale to particles themselves), and thus the diffusion coefficient has units of 1 over time. In order to apply these results to an experimental setup, one needs to rescale them appropriately based on the typical length scales of cells in that experiment (which is also the length scale connected to the rates $a,b$). Generally, for cell size $\Delta x$, we would have the diffusion coefficient $D_{\rm{passive}}=(\Delta x)^2(a+b)/2$ and drift $f=\Delta x (a-b)$. Our choice $\Delta x=1$ is a special case.

We proceeded similarly with Geometry III. Important difference is, that in this geometry it turned out, that $D_{\rm{active}} ^{\rm{nonint.}}=(a-b)^2/(8\alpha)+ D_{\rm{passive}}$. This change is caused by the fact, that in noninteracting case with high Pe, the particles are effectively half of the time stuck in a dead end, and other half of the time propelled in the direction, in which they do not get stuck (we will come back to this later), therefore their mean velocity after averaging over time is $(a-b)/2$, therefore in formula for $D_{\rm{active}} ^{\rm{nonint.}}$ we need to replace $(a-b)^2$ with $(a-b)^2 /4$.

\begin{figure}
    \centering
\begin{subfigure}{.5\textwidth}
  \centering
  \includegraphics[width=\linewidth]{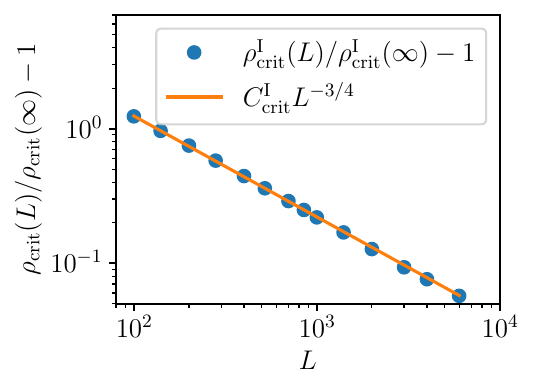}
  \caption{}
\end{subfigure}%
\begin{subfigure}{.5\textwidth}
  \centering
  \includegraphics[width=\linewidth]{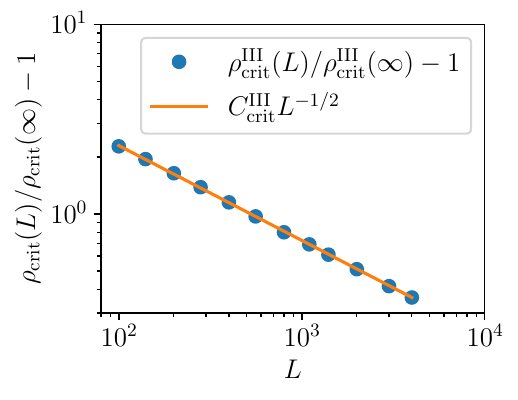}
  \caption{}
\end{subfigure}
    \vskip\baselineskip
\begin{subfigure}{.5\textwidth}
  \centering
  \includegraphics[width=\linewidth]{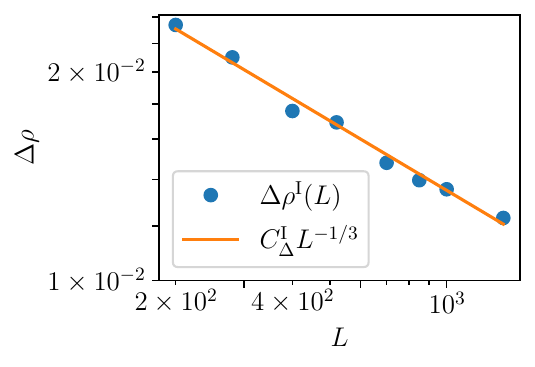}
  \caption{}
\end{subfigure}%
\begin{subfigure}{.5\textwidth}
  \centering
  \includegraphics[width=\linewidth]{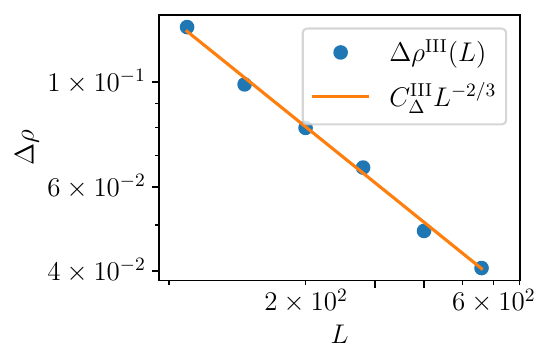}
  \caption{}
\end{subfigure}
\caption{Plots (a) and (b) show how does the critical density $\rho_{\rm{crit}}$ change with increasing the size of the system $L$ for Geometry I and III respectively, with rates $a=1,b=0,\alpha=0.01$, error bars are smaller than the size of the symbols. Values obtained by simulations are in blue {\Large $\bullet$}, and the orange line is a fit by power law. Critical densities for infinite systems obtained as a parameters of this fit are $\rho_{\rm{crit}}^{\rm{I}}(\infty)=0.0761 \pm 0.0001$ and $\rho_{\rm{crit}}^{\rm{III}}(\infty)=0.192 \pm 0.001$, and coefficients are $C_{ \rm{crit}}^{\rm{I}}=2.99\pm0.01$, $C_{ \rm{crit}}^{\rm{III}}=4.38\pm0.02$. Plots (c) and (d) show the width of transition as a function of the size of the system $L$ for the same rates, and  $C_\Delta^{\rm{I}}=0.135\pm0.005$, $C_\Delta^{\rm{III}}=2.75\pm0.04$.}
\label{fig:rho_crit}
\end{figure}

As mean density $\bar \rho$ increases, we can see a sudden drop in $D$. It indicates the transition from the unjammed state to the jammed state. From the data collapse in Fig. \ref{fig:D:ex} we see, that the transition for different sizes of the system $L$ looks the same and can be described just by two numbers, both depending on $L$: its position $\rho_{\rm{crit}}$, and its width $\Delta \rho$. We define the critical density $\rho_{\rm{crit}}$ as the $\bar \rho$ for which $D/D_{\rm{active}} ^{\rm{nonint.}}=1/2$. We define the width of the transition $\Delta \rho$ as the difference between critical density $\rho_{\rm{crit}}$ and the density, where the linear fit of $D$ at $\rho_{\rm{crit}}$ crosses 0. The dependence of $\rho_{\rm{crit}}$ and $\Delta \rho$ on the length of the channel $L$ for both Geometry I and III is in Fig. \ref{fig:rho_crit}. By scaling up of the system, the transition becomes sharper, which should be expected. The critical density is also a decreasing function of length but approaches asymptotically a non-zero limit. In Fig. \ref{fig:rho_crit} (c-d) we see, that $\Delta \rho$ depends on $L$ as a power law, while from Fig. \ref{fig:rho_crit} (a-b) we see, that $\rho_{\rm{crit}}$ approach its nonvanishing asymptote as a power law. Thus the limit for $L \rightarrow \infty$ is $\Delta \rho \rightarrow 0$ and $\rho_{\rm{crit}} \rightarrow \rho_{\rm{crit}} (\infty)>0$. In other words, in the limit of infinite size, there is a dynamical phase transition at some finite density $\rho_{\rm{crit}} (\infty)$. However we stress, that the finite-size effects are significant. These results hold for both Geometry I and III, just with different exponents and coefficients.

\begin{figure}
\centering
\begin{subfigure}{.5\textwidth}
  \centering
  \includegraphics[width=\linewidth]{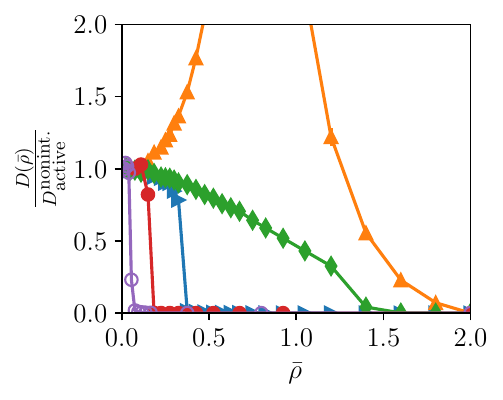}
  \caption{}
\end{subfigure}%
\begin{subfigure}{.5\textwidth}
  \centering
  \includegraphics[width=\linewidth]{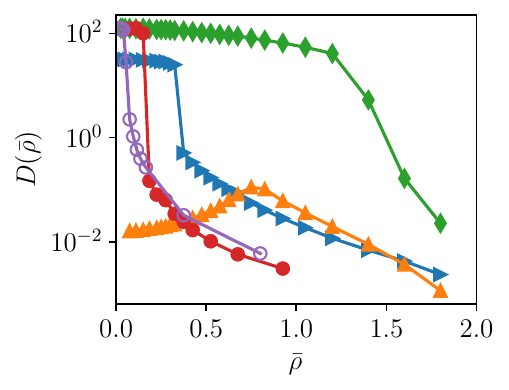}
  \caption{}
\end{subfigure}
\caption{(a) Diffusion coefficient $D$ as a function of the mean density $\bar \rho$ for different geometries for length $L=1000$ and rates $a=1,b=0,\alpha=0.004$. Geometry I is in purple {\Large $\circ$}, Geometry II in red {\Large $\bullet$}, Geometry III in blue $\blacktriangleright$, Geometry IV in orange $\blacktriangle$, and Geometry V in green $\blacklozenge$. The error bars correspond to the standard deviation and, in almost all cases, are smaller than the markers. Data for each geometry are normalised by their own noninteracting limit $D_{ \rm{active}} ^{ \rm{nonint.}}$.
(b) The same data on a semilogarithmic scale, not normalized, with all error bars smaller than the markers.}
\label{fig:D:comp}
\end{figure}

 For each of the geometries described in Fig. \ref{fig:geom}, we calculated the dependence of the diffusion coefficient on density. We can see these diagrams in linear and semilogarithmic scales in Fig. \ref{fig:D:comp}.

The plot shows several things. Geometries III and V sort particles by their orientation, making jamming less likely. In Geometry III, half of the particles are stuck in the pockets. In contrast, in Geometry V, all particles have a cell in the direction of their propulsion, so it is harder for some choking points to develop, suppressing jamming even further. Geometry IV has the lowest diffusion coefficient because it traps particles regardless of their direction of propulsion. The most unique thing about Geometry IV is that the diffusion coefficient is increasing for most of the average densities $\bar \rho$ that are lower than half of the maximum density. That happens because in this interval with increasing density the pockets are filling up and are therefore hindering the movement of the particles less, however, then jamming also kicks in, and the diffusivity starts to decrease after all. One general takeaway is that any deviation from the smooth tube that we tested (smooth tubes being Geometries I and II) pushes the critical density higher.

\section{Density of phases}\label{sec:CGrho}

\begin{figure}
\centering
\begin{subfigure}{.5\textwidth}
  \centering
  \includegraphics[width=\linewidth]{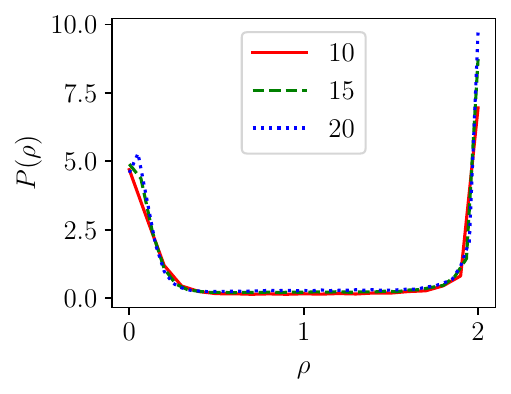}
  \caption{}
\end{subfigure}%
\begin{subfigure}{.5\textwidth}
  \centering
  \includegraphics[width=\linewidth]{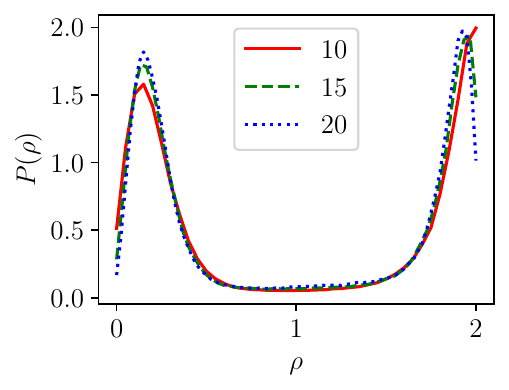}
  \caption{}
\end{subfigure}
\caption{Probability distributions of mean densities for various sizes of coarsegraining window indicated in legend. The size of the system was $L=400$, global mean density $\bar \rho=1$, rates $a=1,b=0,\alpha=0.01$, and we collected data for 2000 different states. In (a) are the results for Geometry I and in (b) for Geometry III.}
\label{fig:CGrhoPdf}
\end{figure}

The jamming is caused by MIPS, i.e. appearance of two phases with higher and lower density. To measure the densities of these phases we use coarse-graining. We choose a length for a coarse-graining window $L_{\rm{CG}}$ and then the mean density in that window is the number of particles inside, divided by the number of cells inside (which is $L_{\rm{CG}}$ for Geometries I and II and $2L_{\rm{CG}}$ for Geometries III-V). We compute this mean density for all the different positions of the coarse-graining window and also we let the system evolve in time and repeat the process in different states of the system. Then we study the distribution of densities obtained by this algorithm. An example distributions for various coarse-graining window sizes in the same run is in Fig. \ref{fig:CGrhoPdf}. In this figure we can see two distinct peaks, one in the region of small densities and another in the region of high densities. These two peaks correspond to the two different phases created by MIPS: gas, where density is low and interaction between particles is thus negligible, and condensate, where density is so high that particles are dynamically frozen in place by their interaction with other particles.

\begin{figure}
    \centering
\begin{subfigure}{.5\textwidth}
  \centering
  \includegraphics[width=\linewidth]{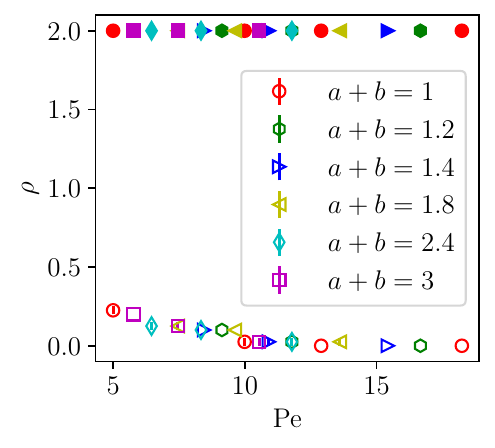}
  \caption{}
\end{subfigure}%
\begin{subfigure}{.5\textwidth}
  \centering
  \includegraphics[width=\linewidth]{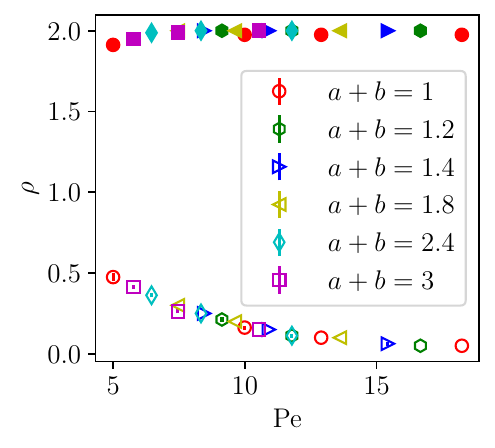}
  \caption{}
\end{subfigure}
    \vskip\baselineskip
\begin{subfigure}{.5\textwidth}
  \centering
  \includegraphics[width=\linewidth]{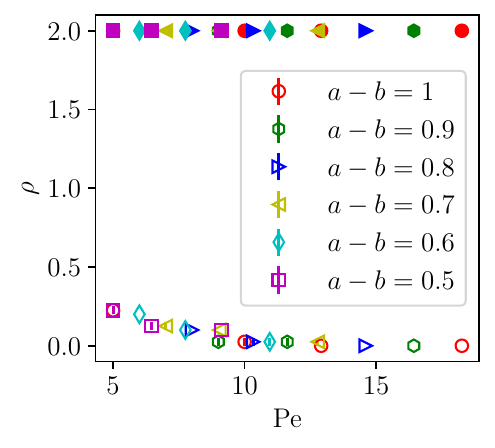}
  \caption{}
\end{subfigure}%
\begin{subfigure}{.5\textwidth}
  \centering
  \includegraphics[width=\linewidth]{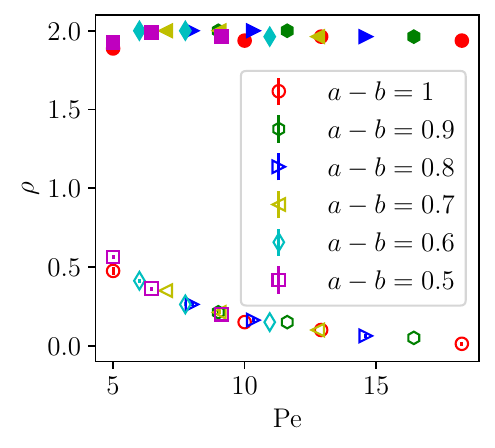}
  \caption{}
\end{subfigure}
\caption{Densities of separated phases as a functions of the Péclet number Pe$=(a-b)/\sqrt{\alpha (a+b)}$. Condensate is in full symbols, and the gas phase is in empty symbols. The length of the system is $L=800$, the global mean density is $\bar \rho=1$, and errors are estimated by comparison of results for various sizes of coarsegraining windows. For plots (a) and (b) (top row), we fix the difference of rates as $a-b=1$ and vary their sum (as indicated in legend) and the rate $\alpha$. In other words, the given symbol has fixed rates $a,b$, and its different data points differ just by rate $\alpha$. For plots (c) and (d) (bottom row), we fix the sum of rates as $a+b=1$ and vary their difference (as indicated in legend) and the rate $\alpha$, analogously to previous plots. Plots (a) and (c) (left side) are for Geometry I, and plots (b) and (d) (right side) are for Geometry III. }
    \label{fig:CGrhoPe}
\end{figure}

As a proxy for the density of separte phases we take the positions of the maxima of the peaks in the distribution. They depend on the parameters of the model only through the Péclet number Pe$=(a-b)/\sqrt{\alpha (a+b)}$, as demonstrated by the data collapse in Fig. \ref{fig:CGrhoPe}. For high Pe the density of the gas phase goes to $0$, while the density of the condensate goes to $k=2$, the maximal density allowed by the exclusion constraint. With lowering of Pe, the density of gas phase grows, because particles more often escape the condensate and therefore there are more particles in the gas phase. At the same time, the most probable density of the condensate stays for quite a long interval of Pe constantly around the $k=2$, just around the Pe=5 it finally starts to decrease. We found it technically difficult to establish the densities pof phases for Peclet number lower than about 5 due to large statistical errors. But the general take-away is, that the densities in the phases depend on the rates only through the Péclet number Pe$=(a-b)/\sqrt{\alpha (a+b)}$. This Péclet number is obtained by comparing the contributions to the diffusion coefficient $D_{\rm{active}} ^{\rm{nonint.}}$. The contribution of active movement is proportional to $(a-b)^2/(2\alpha)$ and the contribution of a passive movement is $D_{\rm{passive}}=(a+b)/2$, so we obtain the Péclet number as the square root of the ratio of these contributions.

\section{Metastable and jammed states}\label{S:JT}

In the rest of this work, we will focus on Geometry III, because of its broken spatial symmetry allowing us to study the ratchet effect.  We call the direction in which the particle can get into a pocket a difficult direction and the opposite direction we call an easy direction. So for example, if there is just 1 particle in the whole system, $b=0$, and the particle is currently propelled in the difficult direction, it will either jump (with the rate $a$) to the next cell along the backbone, or (with the same rate) to the pocket. If it jumps along the backbone, the situation repeats, and if it jumps into a pocket, there is no further cell in its direction and therefore it will stay there, until its direction changes. It is therefore more probable that a particle propelled in the difficult direction will be in the pocket, compared to a particle propelled in the easy direction, in other words, Geometry III tends to sort particles based on their direction of propulsion. Since no direction of propulsion is preferred for the particles themselves, this creates an imbalance in the backbone. This stays to be the case even when the density $\bar \rho$ is high and MIPS occurs, the ratchet effect just gets weaker in such cases, as we will see.

\begin{figure}[ht]\centering
\includegraphics[width=140mm]{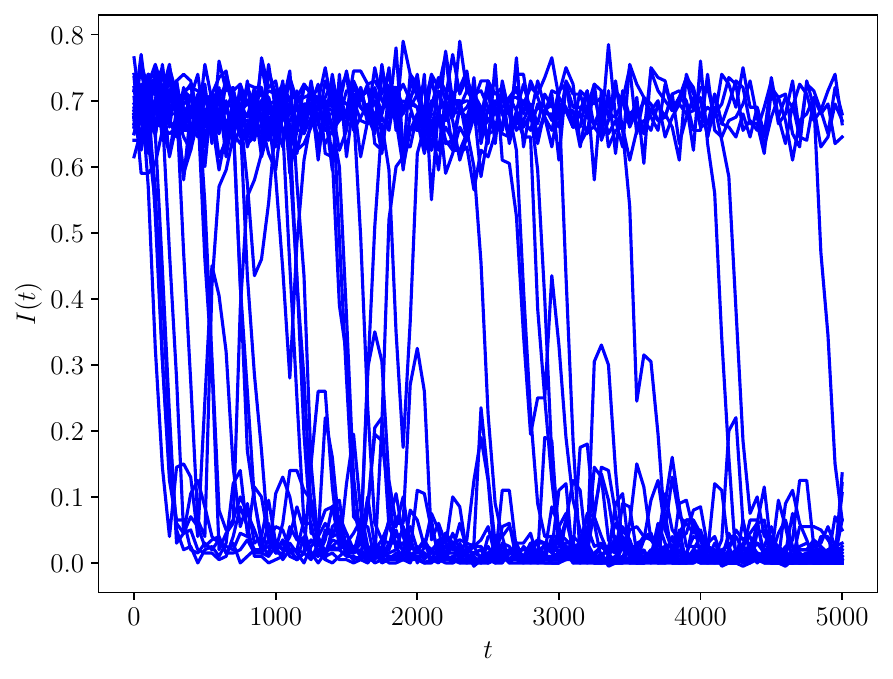}
\caption{Time evolution of ratchet current $I(t)$ for 30 independent runs in a system of Geometry III and length $L=200$, with mean density $\bar \rho =1$, and rates $a=1, b=0, \alpha=0.001$.}
\label{fig:IMS}
\end{figure}

\begin{figure}[ht]\centering
\includegraphics[width=145mm]{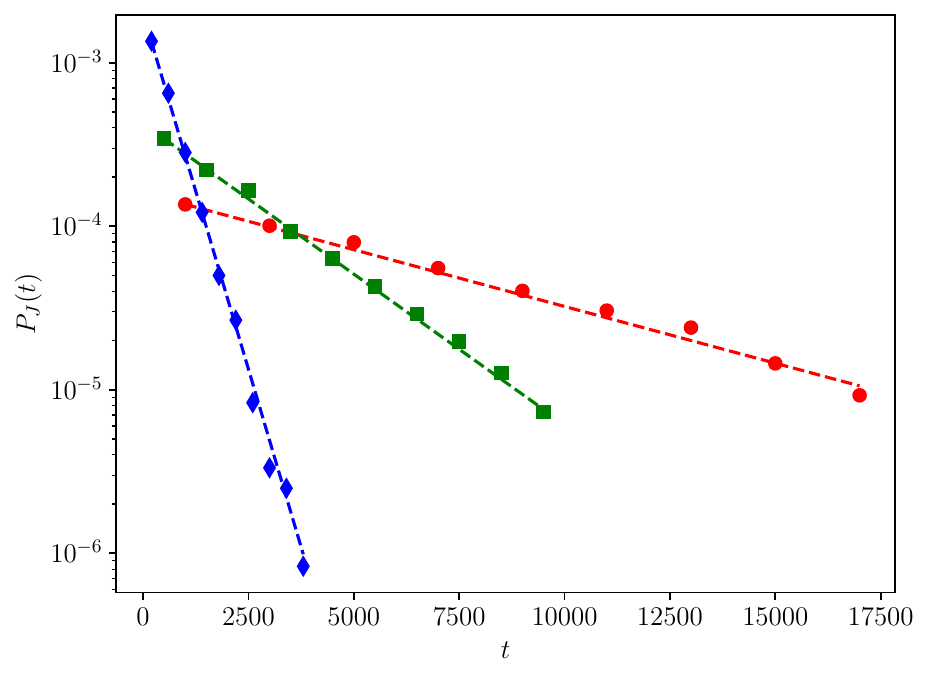}
\caption{Probability density $P_J(t)$ of jamming happening in time $t$ in a system of Geometry III and length $L=200$, with rates $a=1, b=0, \alpha=0.001$ at mean densities $\bar \rho =0.875$ (red {\Large$\bullet$}), $\bar \rho =1$ (green $\blacksquare$), and $\bar \rho =1.25$ (blue $\blacklozenge$). Dashed lines are exponential fits, and errors estimated as standard deviation are smaller than the size of the symbols.}
\label{fig:pdfJ}
\end{figure}

In Fig. \ref{fig:IMS}, we see the time evolution of the ratchet current for a number of independent runs. We initialized the systems in such a way, that all particles propelled in the easy direction were in the backbone and all particles propelled in the difficult direction were in the pockets. We can notice, that for each time evolution, the current at first oscillates around the value 0.7. Then there is at one point in time for each evolution a drop to a much lower current, fluctuating just slightly above 0. After that drop, the current never returns to the original value (and if it did, it would be just a finite-size effect). That indicates that the system experienced a change in the transport regime at that time. More specifically, the system transitions from a metastable state without separated phases to a state with separated phases (dynamically frozen state). In a dynamically frozen state, the ratchet effect still exists, but its magnitude is significantly reduced. Thus, we can use the drop in the ratchet current to determine the time, at which the transition occurred. We shall call it the jamming time. In Fig. \ref{fig:pdfJ} we can see the probability distribution of the jamming times. We can see that the distribution is exponential, indicating that the transition from the metastable unjammed state to the stable jammed state is a Poisson process.

\begin{figure}[ht]\centering
\includegraphics[width=140mm]{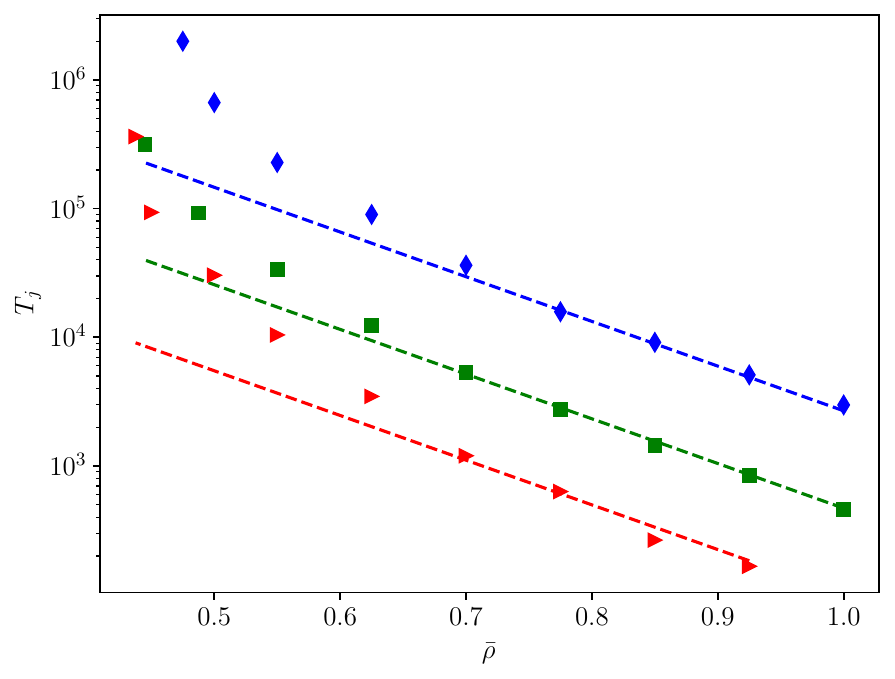}
\caption{The mean jamming time $T_j$ as a function of the mean density $\bar \rho$. The length of the channel is $L=200$, the rates of jumps $a=1,b=0$ are fixed, and the rate of switching directions $\alpha=0.006$ is in red $\blacktriangleright$, $\alpha=0.003$ in green $\blacksquare$, and $\alpha=0.001$ blue $\blacklozenge$. Dashed lines are exponential trends for high enough $\bar \rho$.}
\label{fig:JT}
\end{figure}

Subsequently, we can study the dependence of the mean jamming time it takes the system to freeze $T_j$ on the average density $\bar \rho$. In Fig. \ref{fig:JT} we can see $T_j$ is a decreasing function of the density. The dependence is exponential for higher densities but as densities approach $\rho_{\rm{crit}}$ (as we defined it using the diffusion coefficient, for these parameters we expect the value to be roughly around $\rho_{\rm{crit}} \sim 0.5$), the mean jamming time starts to deviate from this trend, and diverge. The point of view from the perspective of diffusion coefficient fails to capture the fact, that even for $\bar \rho< \rho_{\rm{crit}}$ we still can see clusters. However, those clusters are unstable and decay. Thus, states of the system with separated phases sooner or later transition back into unjammed states, because of finite-size effects. This makes it difficult to distinguish the transitions from the unjammed regime to the jammed regime and back, from current fluctuations within the unjammed regime, therefore, we do not compute $T_j$ for $\bar \rho<0.4$.

The exponential trends in Fig. \ref{fig:JT} have the same slope independent of $\alpha$, but the prefactor is a decreasing function of $\alpha$. In other words, the more often the direction of propulsion changes, the quicker the jammed regime will be reached.

\section{Current-density diagram}\label{S:c-d}

Now we shall investigate the dependence of the ratchet current on the average density. Since there is an interval of densities for which the system can be in two different regimes, as we explained in Section \ref{S:JT}, we calculate both the mean ratchet current in the unjammed state and in the jammed state. 

\subsection{Noninteracting limit}

We start by analytical predictions for unjammed state. If the mean density is so small, that we can neglect all the interactions between the particles, we can calculate the current from one-particle behaviour. If we have just one particle in the whole system, the stationary state can be trivially calculated by setting length $L=1$ and solving the corresponding 4-state Markov chain. We find the steady state of the system as the only eigenvector of the transition matrix with eigenvalue 1. For general rates $a,b,\alpha$ we get probabilities
\begin{eqnarray}
    p_{\rm{E}}^{\rm{P}}=p_{\rm{D}}^{\rm{BB}} &=\frac{b+\alpha}{2 (1+ \frac{b+\alpha}{a+\alpha})(a+\alpha)} \label{eq:ps1p} \\ 
    p_{\rm{D}}^{\rm{P}}=p_{\rm{E}}^{\rm{BB}} &=\frac{1}{2(1+ \frac{b+\alpha}{a+\alpha})} \,, \label{eq:ps1pB} 
\end{eqnarray}
where the lower index denotes orientation of the particle (E: easy, D: difficult), and the upper index its position (P: pocket, BB: backbone).

In the steady state, the flows in and out of the pocket do not contribute to the total current. Then we get the mean current for a single particle in the system of length $L=1$ as
\begin{equation}
    \langle I \rangle = (a-b) p_{\rm{E}}^{\rm{BB}} + (b-a) p_{\rm{D}}^{\rm{BB}} = \frac{(a-b)^2}{2(a+b+2 \alpha)}\,,
\end{equation}
which means, that in our case, when $b=0$, the system has length $L$, and the mean density of particles is $\bar \rho=n/(2 L)$, where $n$ is the total number of particles, the ratchet current in the limit of noninteracting particles should be $\langle I \rangle = \bar \rho a^2 / (a + 2 \alpha)$. In the $a \gg \alpha$ limit this result simplifies to $\langle I \rangle = \bar \rho a$.

This result can be understood in following way: from equations (\ref{eq:ps1p}) and (\ref{eq:ps1pB}) we see, that if we assume $\alpha \ll a$, nearly all particles propelled in the difficult direction are stuck in pockets and nearly all particles propelled in the easy direction are in the backbone creating the current. And since particles themselves have no preference between the 2 directions, and in the geometry, the number of backbone cells is equal to the number of pocket cells, the average density in the backbone is going to be equal to the global average density $\bar \rho$ (this also agrees with (\ref{eq:ps1p}) and (\ref{eq:ps1pB})). If the average density is low enough, so that particles in the backbone are never blocked by other particles in the backbone, the mean current is then $a \bar \rho$. As the density increases, we can can expect the value of the mean current to slowly deviates from the $a \bar \rho$ line, exactly because the backbone contains so many particles, that even though they mostly have the same direction, they can sometimes block each other.

We can also calculate fluctuations of one particle current as
\begin{equation}
    \fl\langle (\sigma I)^2 \rangle =\langle I^2 \rangle -\langle I \rangle^2 = \frac{1}{2}(a-b)^2 -  \frac{(a-b)^4}{4(a+b+2 \alpha)^2}= \frac{1}{4}(a-b)^2 \left(2 - \frac{(a-b)^2}{(a+b+2 \alpha)^2}  \right)  \,,
\end{equation}
which in the case of $n=2 L \bar \rho$ noninteracting particles in the backbone of length $L$ becomes
\begin{equation}
    \langle (\sigma I)^2 \rangle= \bar \rho \frac{1}{2}(a-b)^2 \left(2 - \frac{(a-b)^2}{(a+b+2 \alpha)^2}  \right)  \,.
\end{equation}
We checked that predicted fluctuations match the numerical results.

\subsection{Mean-field approximation}

The correction to interaction between the particles in the unjammed regime can be computed via mean-field. We start by performing a mean-field approximation on Geometry I. That means we assume that all individual cells are uncorrelated. The state of a cell can be described by two integers $n_\rightarrow, n_\leftarrow$ equal to the number of particles propelled in that direction that are in that cell. Because of our exclusion constraint, only states with $n_\rightarrow+ n_\leftarrow \leq k$ are possible; for example, for $k=2$ a cell can be in 6 possible states. We denote probabilities of the cell being in a state with $n_\rightarrow$ and $n_\leftarrow$ particles as $p(n_\rightarrow, n_\leftarrow)$, and our objective is to find an equilibrium probability distribution. Constraints are
\begin{eqnarray}
    1 &=\sum_{n_\rightarrow, n_\leftarrow} p(n_\rightarrow, n_\leftarrow)  \label{eq:MFp1}\\
    \bar \rho &=\sum_{n_\rightarrow, n_\leftarrow}(n_\rightarrow+ n_\leftarrow) p(n_\rightarrow, n_\leftarrow) \label{eq:MFrho}\,,
\end{eqnarray}
the rest of the equations we need to get from dynamics.

We need to find the rates of transitions between the states. For the remainder of this subsection, we assume $k=2$. Some equations may be more general, but most of them are valid only for this case. We then denote $p_n$ the probability of $n$ particles being in a cell, formally $p(0,0)=p_0,p(0,1)+p(1,0)=p_1,p(0,2)+p(1,1)+p(2,0)=p_2$. Then, for a given probability distribution, the time derivative of $p(n_\rightarrow, n_\leftarrow)$ is
\begin{eqnarray}
    \fl \eqalign{ \dot p (n_\rightarrow, n_\leftarrow)=-p (n_\rightarrow, n_\leftarrow) \big( (n_\rightarrow+ n_\leftarrow)(p_0+p_1)(a+b)   +\\
    + [n_\rightarrow+ n_\leftarrow<k](p_1+2 p_2)(a+b) +(n_\rightarrow+ n_\leftarrow)\alpha \big) +}\label{eq:MFdpI1} \\
    \fl+ \big( p (n_\rightarrow+1, n_\leftarrow)(n_\rightarrow+1)+p (n_\rightarrow, n_\leftarrow+1)(n_\leftarrow+1) \big) (p_0+p_1)(a+b) +\label{eq:MFdpI2} \\ 
    \fl+p (n_\rightarrow-1, n_\leftarrow)(p(1,0)+p(1,1)+2p(2,0))(a+b)+\label{eq:MFdpI3} \\ 
    \fl+p (n_\rightarrow, n_\leftarrow-1)(p(0,1)+p(1,1)+2p(0,2))(a+b)+\label{eq:MFdpI4}\\
    \fl+\big(p (n_\rightarrow+1, n_\leftarrow-1)(n_\rightarrow+1)+p (n_\rightarrow-1, n_\leftarrow+1)(n_\leftarrow+1) \big)\alpha \label{eq:MFdpI5} \,,
\end{eqnarray}
where the $[n_\rightarrow+ n_\leftarrow<k]$ is the Iverson bracket, equal to 1 if the condition holds and 0 otherwise, and all $p (n_\rightarrow, n_\leftarrow)$ corresponding to impossible states are trivially 0. The term (\ref{eq:MFdpI1}) is equal to the decrease of $p (n_\rightarrow, n_\leftarrow)$ caused by the state changing to a different one, the terms in brackets correspond to particle jumping away, particle jumping in, and particle switching direction, respectively. The term (\ref{eq:MFdpI2}) is equal to increase of $p (n_\rightarrow, n_\leftarrow)$ caused by particles jumping out of the site, the terms (\ref{eq:MFdpI3}) and (\ref{eq:MFdpI4}) by particles jumping into the site, and term (\ref{eq:MFdpI5}) by particles switching directions. To find the stationary probability distributions, we set $\dot p (n_\rightarrow, n_\leftarrow)=0$ for all possible states. In combination with constraints (\ref{eq:MFp1}) and (\ref{eq:MFrho}), we have the full system of equations, this system has 2 more equations than unknowns, but the equations are linearly dependent.

Let us now solve this system of equations. We can use the symmetry to get $p(0,1)=p(1,0)=p_1/2$. Then from the equation $\dot p (0,0)=0$ we get
\begin{equation}
    0=-p_0(a+b)(p_1+2p_2) + p_1 (a+b)(p_0+p_1) \rightarrow 2 p_0 p_2 = p_1^2 \,,
\end{equation}
combining this with (\ref{eq:MFp1}) and (\ref{eq:MFrho}) and solving for $p_0,p_1,p_2$, we get
\begin{eqnarray}
    p_0 &= 1+ \frac{1-\sqrt{1+\bar \rho(2-\bar \rho)}}{2} - \frac{\bar \rho}{2} \\
    p_1 &=-1+\sqrt{1+\bar \rho(2-\bar \rho)}\\
    p_2 &= \frac{\bar \rho}{2}+ \frac{1-\sqrt{1+\bar \rho(2-\bar \rho)}}{2} \,.
\end{eqnarray}
It remains to be determined how is $p_2$ distributed among $p(0,2),p(1,1),p(2,0)$. From symmetry $p(0,2)=p(2,0)$, and from $\dot p (0,2)=0$ we get
\begin{equation}
    \fl0=-p(0,2) \big(2 (a+b)(p_0+p_1)+ 2\alpha \big)+p(0,1)(a+b)(p(0,1)+p(1,1)+2p(0,2))+p(1,1)\alpha \,,
\end{equation}
using previous results we can get
\begin{equation}
    p(0,2)=p(2,0)=\frac{p(1,1)}{2}=\frac{p_2}{4} \,,
\end{equation}
so the probability distribution is
\begin{eqnarray}
    p(0,0) &=1+ \frac{1-\sqrt{1+\bar \rho(2-\bar \rho)}}{2} - \frac{\bar \rho}{2} \\
    p(0,1)=p(1,0) &= \frac{-1+\sqrt{1+\bar \rho(2-\bar \rho)}}{2} \\
    p(0,2)=p(2,0) &=  \frac{\bar \rho+1-\sqrt{ 1+\bar \rho(2-\bar \rho)}}{8}\\
    p(1,1) &=  \frac{\bar \rho+1-\sqrt{ 1+\bar \rho(2-\bar \rho)}}{4}\,.
\end{eqnarray}
This can be generalized as double Poisson distribution
\begin{equation}
    p (n_\rightarrow, n_\leftarrow)=\frac{1}{\mathcal{N}} \frac{\lambda_\rightarrow^{n_\rightarrow}\lambda_\leftarrow^{n_\leftarrow}}{n_\rightarrow! n_\leftarrow!}\,,
\end{equation}
for possible states and 0 otherwise, where thanks to symmetry $\lambda_\rightarrow= \lambda_\leftarrow \equiv \lambda$, where
\begin{equation}
    \lambda(\bar \rho)= \frac{\sqrt{ 1+\bar \rho(2-\bar \rho)} +\bar \rho -1}{2(2-\bar \rho)} \,,
\end{equation}
 and $\mathcal{N}=1+2\lambda+2\lambda^2$ is a normalization factor. Notice, that those probabilities do not depend on the rates, this will not be the case in Geometry III.

\begin{figure}[ht]\centering
\includegraphics[width=140mm]{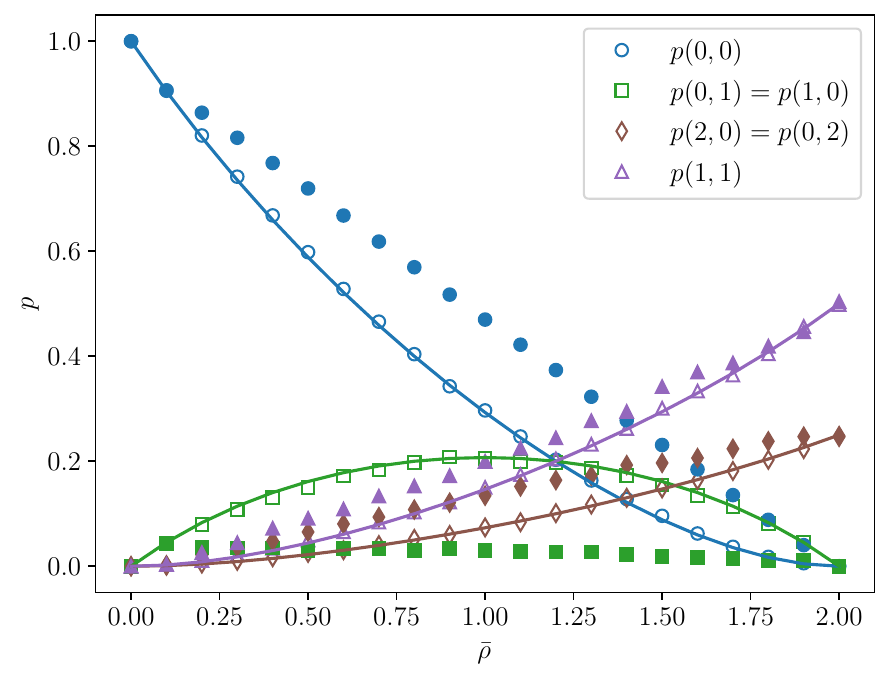}
\caption{Probabilities of a cell in Geometry I being in a given state as functions of mean density $\bar \rho$. States are indicated in the legend, full symbols are from simulation for rates $a=1,b=0,\alpha=0.01$ (thus Pe$=10$), empty symbols for rates $a=1,b=0,\alpha=4$ (Pe$=0.5$), solid lines are analytical results of the mean field calculations.}
\label{fig:MFp}
\end{figure}

When Pe is low, the probabilities obtained from simulations are in good agreement with these analytical formulas, as we show in Fig. \ref{fig:MFp}. However in simulations with high Pe, the agreement only lasts while $\bar \rho <\rho_{\rm{crit}}^{\rm{I}} $, for higher $\bar \rho$ the MIPS occurs, the correlations between the states of neighbouring cells become big, and mean-field approximation does not give reliable predictions. These data are also shown in Fig. \ref{fig:MFp}.

Now let us apply this approach to Geometry III. We imagine the whole system as composed of individual components, each comprising one backbone cell and one pocket cell attached to it, and as a mean-field approximation we assume that these individual components are uncorrelated.  The state of an individual component can be described by four integers $n^{\rm{BB}}_\rightarrow, n^{\rm{BB}}_\leftarrow,n^{\rm{P}}_\rightarrow, n^{\rm{P}}_\leftarrow$ equal to the number of particles propelled in that direction that are in that component in backbone cell, or pocket cell respectively. Because of our exclusion constraint, only states with $n^{\rm{I}}_\rightarrow+ n^{\rm{I}}_\leftarrow \leq k$ (where I is either BB or P) are possible. In case of $k=2$, the individual component can be in 36 different states. The constraints are then
\begin{eqnarray}
    1 &= \sum_{n^{\rm{BB}}_\rightarrow, n^{\rm{BB}}_\leftarrow,n^{\rm{P}}_\rightarrow, n^{\rm{P}}_\leftarrow} p(n^{\rm{BB}}_\rightarrow, n^{\rm{BB}}_\leftarrow,n^{\rm{P}}_\rightarrow, n^{\rm{P}}_\leftarrow)  \label{eq:MFp1III}\\
    2 \bar \rho &= \sum_{n^{\rm{BB}}_\rightarrow, n^{\rm{BB}}_\leftarrow,n^{\rm{P}}_\rightarrow, n^{\rm{P}}_\leftarrow}(n^{\rm{BB}}_\rightarrow+ n^{\rm{BB}}_\leftarrow+n^{\rm{P}}_\rightarrow+ n^{\rm{P}}_\leftarrow) p(n^{\rm{BB}}_\rightarrow, n^{\rm{BB}}_\leftarrow,n^{\rm{P}}_\rightarrow, n^{\rm{P}}_\leftarrow) \label{eq:MFrhoIII}\,.
\end{eqnarray}

Unlike in the Geometry I, the directions are not symmetric, let us assume, that $\rightarrow$ is the difficult direction, and $\leftarrow$ is the easy one. Finally, for simplicity, we assume that $b=0$, and we still consider specifically the case $k=2$. Now we can write the time derivatives of probabilities
\begin{eqnarray}
    \fl\eqalign{\dot p(n^{\rm{BB}}_\rightarrow, n^{\rm{BB}}_\leftarrow,n^{\rm{P}}_\rightarrow, n^{\rm{P}}_\leftarrow)=- p(n^{\rm{BB}}_\rightarrow, n^{\rm{BB}}_\leftarrow,n^{\rm{P}}_\rightarrow, n^{\rm{P}}_\leftarrow) \big( n^{\rm{BB}}_\rightarrow [n^{\rm{P}}_\rightarrow+ n^{\rm{P}}_\leftarrow<k] a + \\
    + n^{\rm{P}}_\leftarrow [n^{\rm{BB}}_\rightarrow+ n^{\rm{BB}}_\leftarrow<k]a + (n^{\rm{BB}}_\rightarrow+ n^{\rm{BB}}_\leftarrow) p_{\rm{free}} a + [n^{\rm{BB}}_\rightarrow+ n^{\rm{BB}}_\leftarrow<k] (\lambda_\rightarrow+\lambda_\leftarrow) a +  \\
    + (n^{\rm{BB}}_\rightarrow+ n^{\rm{BB}}_\leftarrow+n^{\rm{P}}_\rightarrow+ n^{\rm{P}}_\leftarrow)\alpha \big)+ \label{eq:MFdpIII1}} \\
   \fl +\big( p(n^{\rm{BB}}_\rightarrow+1, n^{\rm{BB}}_\leftarrow,n^{\rm{P}}_\rightarrow-1, n^{\rm{P}}_\leftarrow) (n^{\rm{BB}}_\rightarrow+1)+ p(n^{\rm{BB}}_\rightarrow, n^{\rm{BB}}_\leftarrow-1,n^{\rm{P}}_\rightarrow, n^{\rm{P}}_\leftarrow+1) (n^{\rm{P}}_\leftarrow+1)\big)a \label{eq:MFdpIII2} \\
  \fl  + \big(p(n^{\rm{BB}}_\rightarrow+1, n^{\rm{BB}}_\leftarrow,n^{\rm{P}}_\rightarrow, n^{\rm{P}}_\leftarrow) (n^{\rm{BB}}_\rightarrow+1)+p(n^{\rm{BB}}_\rightarrow, n^{\rm{BB}}_\leftarrow+1,n^{\rm{P}}_\rightarrow, n^{\rm{P}}_\leftarrow) (n^{\rm{BB}}_\leftarrow+1) \big) p_{\rm{free}}a + \label{eq:MFdpIII3} \\
  \fl  + \big(p(n^{\rm{BB}}_\rightarrow-1, n^{\rm{BB}}_\leftarrow,n^{\rm{P}}_\rightarrow, n^{\rm{P}}_\leftarrow) \lambda_\rightarrow+p(n^{\rm{BB}}_\rightarrow, n^{\rm{BB}}_\leftarrow-1,n^{\rm{P}}_\rightarrow, n^{\rm{P}}_\leftarrow) \lambda_\leftarrow \big)a + \label{eq:MFdpIII4} \\
 \fl  \eqalign{ +\big( p(n^{\rm{BB}}_\rightarrow+1, n^{\rm{BB}}_\leftarrow-1,n^{\rm{P}}_\rightarrow, n^{\rm{P}}_\leftarrow) (n^{\rm{BB}}_\rightarrow+1)+ p(n^{\rm{BB}}_\rightarrow-1, n^{\rm{BB}}_\leftarrow+1,n^{\rm{P}}_\rightarrow, n^{\rm{P}}_\leftarrow) (n^{\rm{BB}}_\leftarrow+1)+ \\
  +p(n^{\rm{BB}}_\rightarrow, n^{\rm{BB}}_\leftarrow,n^{\rm{P}}_\rightarrow+1, n^{\rm{P}}_\leftarrow-1) (n^{\rm{P}}_\rightarrow+1)+p(n^{\rm{BB}}_\rightarrow, n^{\rm{BB}}_\leftarrow,n^{\rm{P}}_\rightarrow-1, n^{\rm{P}}_\leftarrow+1) (n^{\rm{P}}_\leftarrow+1)\big)\alpha \,, }\label{eq:MFdpIII5}
\end{eqnarray}
where we denoted
\begin{equation}
    p_{\rm{free}}=\sum_{n^{\rm{BB}}_\rightarrow, n^{\rm{BB}}_\leftarrow,n^{\rm{P}}_\rightarrow, n^{\rm{P}}_\leftarrow} [n^{\rm{BB}}_\rightarrow+ n^{\rm{BB}}_\leftarrow<k]  p(n^{\rm{BB}}_\rightarrow, n^{\rm{BB}}_\leftarrow,n^{\rm{P}}_\rightarrow, n^{\rm{P}}_\leftarrow)
\end{equation}
the probability, that in the backbone cell is less than $k$ particles, and
\begin{equation}
    \rho_\leftrightarrows=\sum_{n^{\rm{BB}}_\rightarrow, n^{\rm{BB}}_\leftarrow,n^{\rm{P}}_\rightarrow, n^{\rm{P}}_\leftarrow} n^{\rm{BB}}_\leftrightarrows  p(n^{\rm{BB}}_\rightarrow, n^{\rm{BB}}_\leftarrow,n^{\rm{P}}_\rightarrow, n^{\rm{P}}_\leftarrow)
\end{equation}
the mean numbers of particles of indicated orientation in the backbone cell. The term (\ref{eq:MFdpIII1}) is equal to the decrease of $p(n^{\rm{BB}}_\rightarrow, n^{\rm{BB}}_\leftarrow,n^{\rm{P}}_\rightarrow, n^{\rm{P}}_\leftarrow)$ caused by the state changing to a different one, the terms in brackets correspond to particle jumping from backbone cell to pocket, from pocket cell to backbone, away along the backbone, jumping in from neighbouring component, and particle switching direction, respectively. The term (\ref{eq:MFdpIII2}) is equal to increase of $p(n^{\rm{BB}}_\rightarrow, n^{\rm{BB}}_\leftarrow,n^{\rm{P}}_\rightarrow, n^{\rm{P}}_\leftarrow)$ caused by particles jumping from backbone cell to pocket and vice versa, the term (\ref{eq:MFdpIII3}) by particles jumping out of the component along the backbone, the term (\ref{eq:MFdpIII4}) by particles jumping into the component, and term (\ref{eq:MFdpIII5}) by particles switching directions. 

It is unlikely, that the resulting system of equations has an analytical solution. Therefore, we solve the set of equations numerically, by iteration until convergence, where for a given $\bar \rho$ we start from a guess consistent with constraints (\ref{eq:MFp1III}) (\ref{eq:MFrhoIII}) and then iteratively evolve it in time according to (\ref{eq:MFdpIII1}-\ref{eq:MFdpIII5}) until we reach a stationary probability distribution. Once we have this probability distribution, we can calculate the corresponding mean ratchet current as $\langle I \rangle=(\rho_\leftarrow-\rho_\rightarrow)p_{\rm{free}}$.

\subsection{Jammed regime}

The jammed current for high densities should decrease linearly with average density because the value of this current is proportional to the size of the unjammed part of the backbone. And since almost all particles are in the jammed cluster the size of that cluster is going to be roughly $n/(2 k)$, where $n$ is the number of particles, and the 2 is there because in unit length, there are two cells: one in the backbone and the second is a pocket. Then the part not occupied by them is equal to $L-n/(2 k)$. Since we are interested in the mean current, we divide the length of the free region of the backbone by $L$ and get the linearly decreasing factor $1-\bar \rho / k$. When the particle at the edge of the cluster changes its direction, it departs from the cluster. If its new direction is the difficult one, it will most likely end up in a pocket. However, if its new direction is the easy direction, it will most likely travel the full free region of the backbone. There are two effects we have not considered so far, both stemming from the fact, that the free region of the backbone is not empty. Because of that, the cluster will be, on average, shorter because it will consist of a smaller number of particles than we assumed. On the other hand, since the particles in the free region are almost all traveling in the easy direction, the end of the cluster is moving toward those particles, as the particles in front of them arrive at the end. Therefore, the particles will not cover the whole free region of the backbone that was there, when they escaped the cluster.

\begin{figure}[ht!]\centering
\includegraphics[width=140mm]{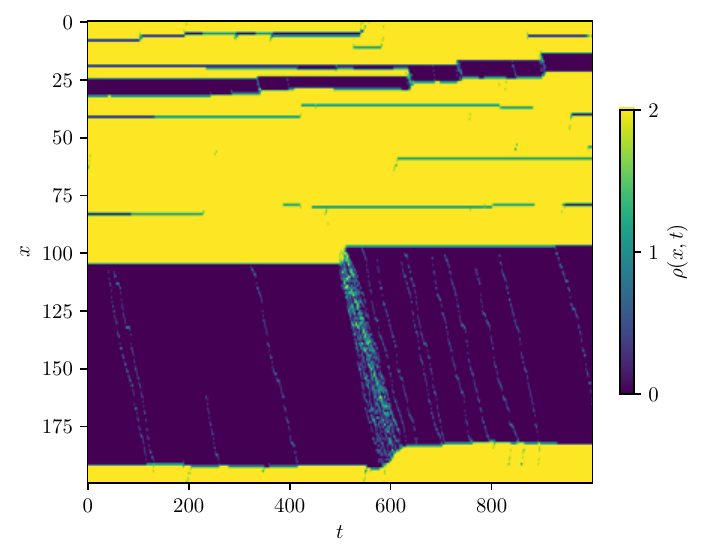}
\caption{An example time evolution in system of length $L=200$ with rates $a=1,b=0,\alpha=0.001$. The plot shows the number of particles in a time and place in the backbone. We do not show the pockets. In this example, we can see a chain reaction of many particles going through the free region of the backbone after one particle at the end of the cluster changes its direction. We can also see that inside the cluster, there are holes.}
\label{fig:example}
\end{figure}

We argue that these two effects cancel each other since if particles did not overtake each other, they would travel the part of the backbone corresponding to a free region in case of all other particles being in the cluster. Particles can overtake, but that means, that some will travel longer parts and some shorter parts, and the differences caused by overtaking will average out to 0. Therefore, the resulting current will be $1-\bar \rho /k$ times the rate of particles escaping the side of the cluster that is in the easy direction to the rest of the cluster. The rate is at least $k \alpha$ since $k$ particles are in the last cell (so not blocked by other particles) and they all have the rate of changing direction $\alpha$. However, the rate can be much larger, that is because after one of the $k$ particles at the end of the cluster changes its direction, it does not only go on its own, but it also stops blocking other particles with the right orientation so they can escape as well. An example of such a chain reaction is in the Fig. \ref{fig:example}.

From Fig. \ref{fig:example} we also see, that our explanation so far was incomplete in the sense that we assumed that in the backbone there is just one cluster and one free region, while in Fig. \ref{fig:example} we see, that it does not have to be the case. However, it changes nothing about the fact, that the free regions of the backbone make up $1-\bar \rho /k$ of the whole backbone. Therefore, if the rate of particles escaping the side of the cluster does not depend on the length of the cluster (which we assumed already, because otherwise it would have been different for different $\bar \rho$), our considerations should hold even in the case of more clusters. The only problem is, that in the case of small free regions particles propelled in the difficult direction might have a non-negligible probability of crossing such a free region. However, since on average they travel just 2 cells, this contribution is still negligible.

\subsection{Comparisson with simulation results}

\begin{figure}[ht!]\centering
\includegraphics[width=140mm]{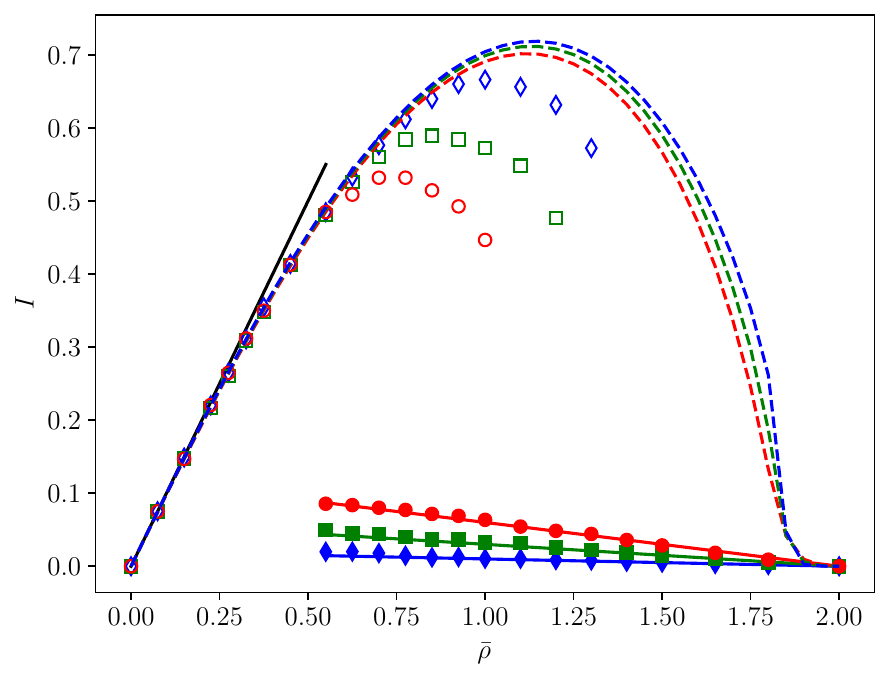}
\caption{Relation between current $I$ and average density $\bar \rho$. The length of the system is $L=200$, and the rates of jumps are fixed as $a=1,b=0$. In red {\Large$\circ$},{\Large$\bullet$}, are data for $\alpha=0.006$, in green $\square$,$\blacksquare$ for $\alpha=0.003$, and in blue $\lozenge$,$\blacklozenge$ for $\alpha=0.001$. The values of the current in the unjammed regime are in the empty {\Large$\circ$},$\square$,$\lozenge$, and the values of the current in the jammed regime are in the filled {\Large$\bullet$},$\blacksquare$,$\blacklozenge$. The black line $a \bar \rho$ is the limit for noninteracting particles, colored dashed lines are results of mean-field calculations, and solid colored lines are fits for jammed regime $K \alpha a (1-\bar \rho /k)$, where $K$ is empirically $\simeq 20$.}
\label{fig:c-d}
\end{figure}

We simulate the system in both regimes to get both the jammed current and unjammed current. For very small densities, we do not see MIPS and therefore get only the unjammed current. Then there is an interval of $\bar \rho$ for which both regimes are possible, but unjammed regime is metastable. For such densities we get the value of unjammed current by averaging over times before jamming occurred and the value of jammed current by averaging over times after that. For high densities, it is hard to properly compute the value of the unjammed current because the jamming time gets so small that it is shorter than the transient time after the initialization. In other words, at the beginning of the simulation, we get behavior that reflects more our way of initializing the system than actual physics. This fact is not always problematic since we can wait some time before collecting the data. And as we saw in Fig. \ref{fig:pdfJ}, the jamming is a Poisson process, therefore the specific choice of time we let the system equilibrate does not matter as long as it removes the initial transient, since after that the system is just in one of the two regimes we described, the probabilities of the system being in one of those regimes do change with time, however, the regimes themselves do not. The initial transient state only makes it impossible for us to compute unjammed currents for high densities because for those, the system jams itself sooner than it starts to give reliable unjammed currents. The results are shown in Fig. \ref{fig:c-d}. We can see for each $\alpha$ the two branches, the branch of unjammed current starts at $\bar \rho=0$, initially grows linearly, and eventually curves down, the branch of jammed currents lies below the unjammed branch, it starts at finite density and linearly decreases all the way to zero current at $\bar \rho=k$.

We can see that for very small densities the straight line calculated from the noninteracting limit indeed match the simulation. But as we expected, after the particles start to get in each others way, the mean-field approximation gives more reliable results than the noninteracting limit. However, once $\bar \rho$ reaches values for which the jamming is possible the mean-field approximation also departs from the numerical results. This makes sense because jamming naturally implies strong correlations and therefore the breakdown of the mean-field approximation. Despite that, the mean-field correctly predicts that ratchet current decreases with increasing $\alpha$, it just greatly underestimates the difference.

Our predictions regarding the current of the jammed regime seem to hold along the whole interval, for which we could simulate the jammed current. From our considerations, we were not able to get the rate of particles escaping the cluster, we only established a lower bound $k\alpha$, and alluded that we expect the value to be much larger than this lower bound. From the simulations we got, that for $k=2$ the rate is roughly $20 \alpha$, which is consistent with our expectations.


\section{Comparison with hydrodynamic approximation}\label{sec:hydro}

Assuming sufficiently smooth spatial variantions of particle density
and neglecting long-range correlations, it is possible to derive
coarse-grained description of the model in terms of hydrodynamic
equations, which in this case assume the form of coupled non-linear
diffusion equations. Following the procedure described in
Ref. \cite{sla_kot_23} which itself draws the idea from
Ref. \cite{ari_kra_mal_14}, we can introduce the fugacities $\lambda_\rightarrow$ and
$\lambda_\leftarrow$ corresponding to particles moving to the right and to the
left, respectively. They are related to the corresponding particle
densities by
\begin{equation}
\rho_\leftrightarrows=\psi_k(\lambda_\rightarrow+\lambda_\leftarrow)\lambda_\leftrightarrows \,,
  \end{equation}
where the function $\psi_k(\lambda_c)$ takes into account the exclusion
constraint imposing at most $k$ particles on a single site. For our case
$k=2$ we have
\begin{equation}
  \psi_2(\lambda_c)=\frac{1+\lambda_c}{1+\lambda_c+\frac{1}{2}\lambda_c^2}\, .
\end{equation}
For convenience, we introduce the total and differential quantities
\begin{equation}
    \eqalign{\lambda_c &= \lambda_\rightarrow + \lambda_\leftarrow\\
    \lambda_d &= \lambda_\rightarrow - \lambda_\leftarrow\\
    \rho_c &= \rho_\rightarrow + \rho_\leftarrow\\
    \rho_d &= \rho_\rightarrow - \rho_\leftarrow \,.
    }
\end{equation}
In terms of the fugacities, the hydrodynamic equations have the form
\begin{equation}
    \eqalign{
    \frac{\partial}{\partial   t}\psi_2(\lambda_c)\lambda_c &=
    -\frac{\partial}{\partial x} j_c\\
    \frac{\partial}{\partial   t}\psi_2(\lambda_c)\lambda_d &=
    -\frac{\partial}{\partial x} j_d - 2\alpha\psi_2(\lambda_c)\lambda_d\,,
    }
\end{equation}
where the currents are
\begin{equation}
    \eqalign{j_c &= \big(\psi_2(\lambda_c)\big)^2\Big(-D_0\frac{\partial}{\partial x}\lambda_c+f_0\lambda_d\Big)           \\
    j_d &= \big(\psi_2(\lambda_c)\big)^2\Big(-D_0\frac{\partial}{\partial x}\lambda_d+f_0\lambda_c\Big)   \,.}
\end{equation}
The parameters $D_0$ and $f_0$ are the diffusion coefficient and
active drift for non-interacting particles, respectively.
For further details see \cite{sla_kot_23}.

We solved these equations in stationary (time-independent) state in
piecewise-one-dimensional geometry, where on a segment of the backbone
of length $L_1$ is attached a segment of the pocket, of the same
length $L_1$. The segment of the backbone is endowed by periodic
boundary conditions, while the dead end of the pocket corresponds to
reflective boundary conditions and the point of the joint, i.e. where
the end of the pocket is joined to the backbone, must satisfy the
conservation of currents. The periodic boundary conditions on the
backbone imply that the solution exhibiting the jamming is
excluded. This must be taken into account when comparing the results
to simulations. 

\begin{figure}[ht!]
  \includegraphics[scale=0.44]{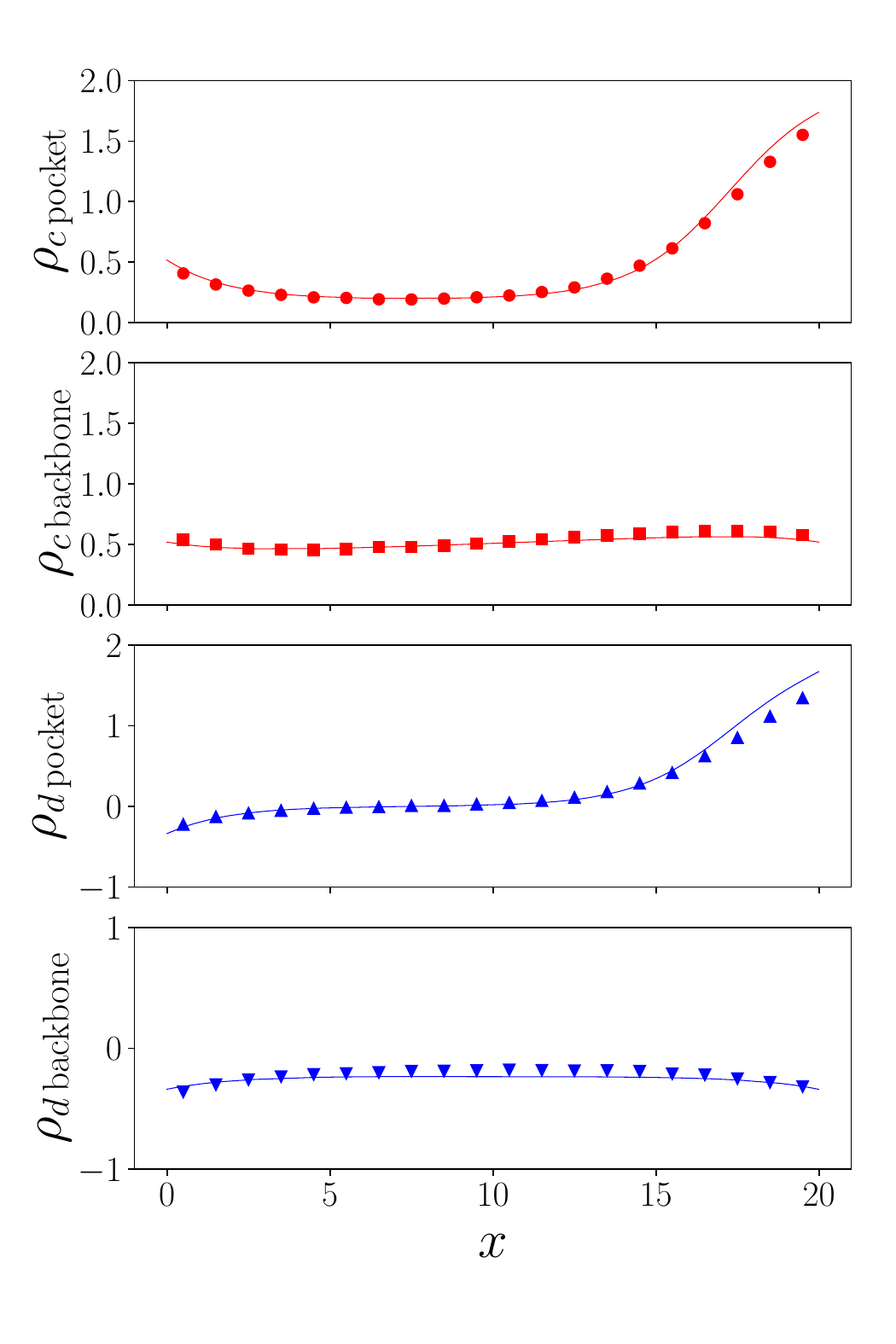}
  \includegraphics[scale=0.44]{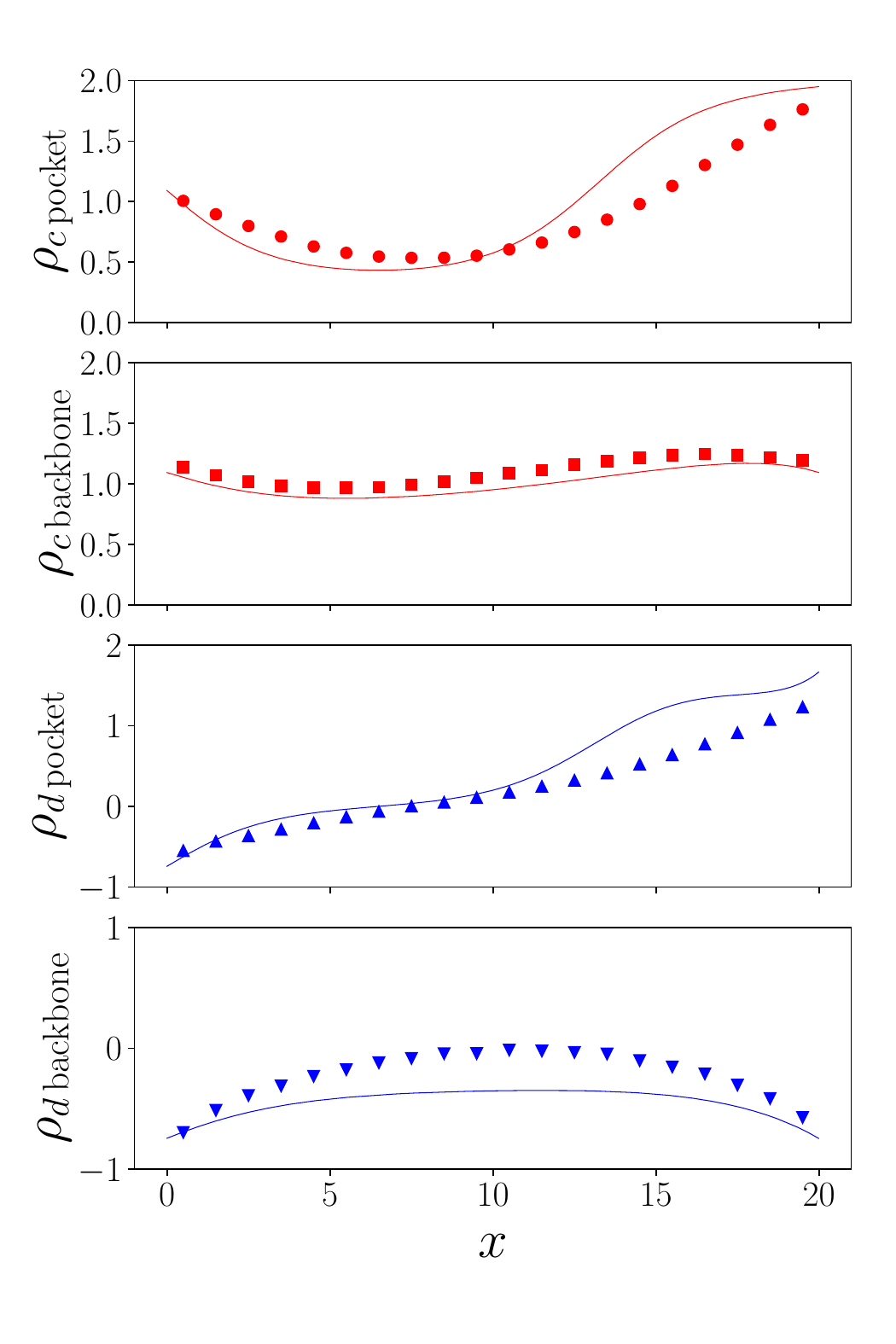}
  \caption{Average density of particles on the segment
    of the backbone and on the adjacent pocket. The numerical solution
    of the hydrodynamic equation is shown by the solid line, and the results of
    simulations are shown by symbols. The length of the segment
    is $L_1=20$. In the left panel, the average
    density is $\overline{\rho}_c=0.5$, in the right panel it is $\overline{\rho}_c=1$. The
    length of the pocket is $L_1=20$. The parameters of the hydrodynamic equations are $D_0=1$,
    $f_0=0.6$, $\alpha=0.01$, which correspond to the parameters used in
    simulations $a=1.3$, $b=0.7$, $\alpha=0.01$.
  }
  \label{hydro-rho}
\end{figure}

We show in Fig.  \ref{hydro-rho} the densities $\rho_c$ and $\rho_d$
on the backbone segment and on the pocket segment. We compare the
results of hydrodynamic approximations with simulations for average
densities $\overline{\rho}_c=0.5$ and $\overline{\rho}_c=1$. The
qualitative agreement is very good, quantitatively they agree very
well at smaller densities but there are
deviations where the density approaches its maximum value of $2$,
which occurs especially at the dead end of the pocket.

\begin{figure}[ht!]
  \includegraphics[width=140mm]{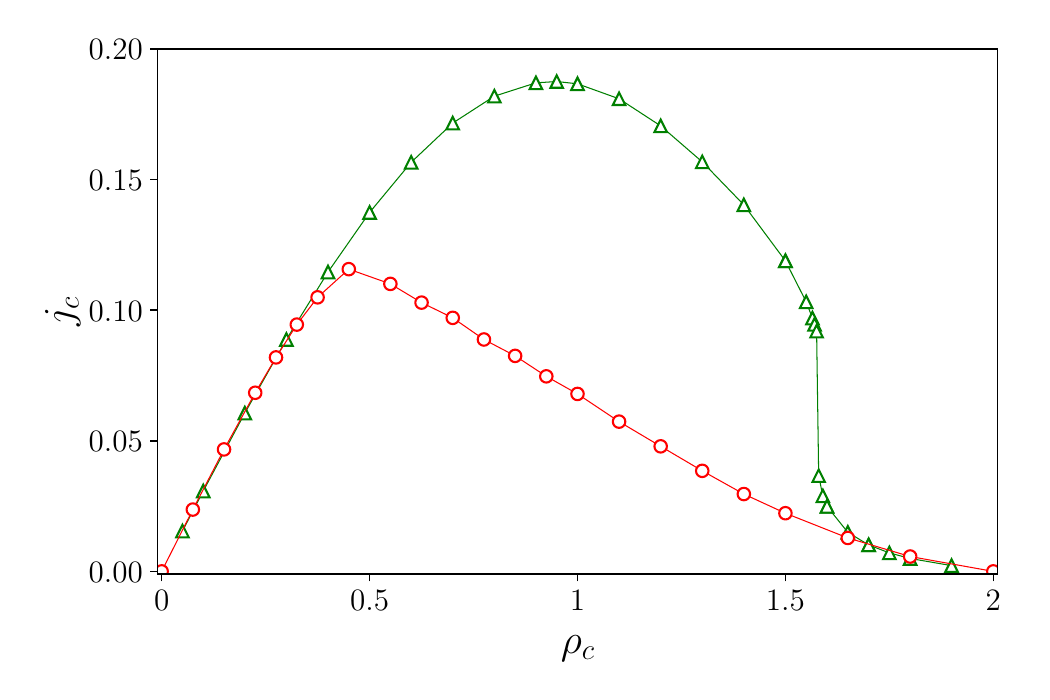}
  \caption{Stationary current in the system with pockets of length
    $L_1=20$. We compare the  
    results of the numerical solution of hydrodynamic equations (symbols
    $\triangle$)  with results of simulations (symbols
    {\Large $\circ$}). The parameters of the hydrodynamic equations are $D_0=1$,
    $f_0=0.6$, $\alpha=0.01$, which correspond to the parameters used in
    simulations $a=1.3$, $b=0.7$, $\alpha=0.01$. In the solution of
    hydrodynamic equations, strict spatial periodicity is enforced,
    thus excluding the jamming.
  }
  \label{hydro-current}
\end{figure}

We also calculated the stationary ratchet current. The results are
shown in Fig. \ref{hydro-current} together with the data from
simulations. We can clearly see that the 
hydrodynamic approximation agrees very well  with simulations up to a
density $\rho_{c}<\rho_{c1}$, where in our case $\rho_{c1}\simeq
0.5$. Beyond this density, the current in the simulations is
suppressed due to jamming. On the contrary, the jamming is
artificially excluded by imposing periodic boundary conditions in the
hydrodynamic approximation, as already mentioned above. So, for
$\rho_{c}>\rho_{c1}$ the hydrodynamic results must be interpreted as
corresponding to a 
hypothetical current that would be observed if the jamming does not
happen. This is exactly the metastable current seen in simulations, in
the range of 
densities where it is observable.

\section{Conclusion}\label{sec:conclusion}

We investigated a lattice model of active particles in a quasi-1D
plain or periodically modulated channel.
We investigted five different geometries of the channel, namely a
plain one, which has two variants differing in the cell capacity, and
then geometry with periodic one-side pockets, with periodic alternating
pockets and a tilted ladder.
 The
steric interaction between particles  was
accounted for by an exclusion constraint on the number of particles on
a single site. The Markovian dynamics is governed by rates of jumps
between neighboring sites. From this point of view the model is a
generalization of the ASEP model. The activity of the particles was represented by the
fact, that each particle had its direction of propulsion, in which the
particle was jumping with the rate higher than in the opposite
direction, causing it to drift in that direction. Particles switch
their directions at a fixed rate. It turns out that the behavior of
the model is determined by a single combination of all three rates,
(i.e. rates of jumps forward, jumps backward and switching rate) which
is the P\'eclet number. Here we were mainly interested in the regime
of high P\'eclet number.
The phenomena we investigated were tracer diffusion, ratchet current
and jamming due to motility-induced phase separation.

Besides direct numerical simulations, we investigated the model also
analytically in the one-particle limit and in mean-field
approximation. We also compared the simulation results with numerical
solution of corresponding hydrodynamic equations.  

As a first step, we studied the tracer diffusion coefficient. For low
densities we get results consistent with the limit of noninteracting
particles. When the mean density is increased, the diffusion
coefficient drops at some point, indicating a transition from the
regime without separated phases to the regime with motility-induced
phase separation, in which clusters of particles form and hinder the
motion of those particles, reducing the diffusion coefficient by
orders of magnitude. The data collapse on an universal curve suggests
that the transition between these 
regimes can be described just by two parameters: its position
$\rho_{crit}$ and its width $\Delta \rho$. We studied the behaviour of
these parameters when the size of the system is increased. We found
 that while the width $\Delta \rho$ goes asymptotically to 0, the
critical density $\rho_{crit}$ has a non-zero asymptote, which implies
that sharp transition occurs at finite average density in the
thermodynamic limit. 

The behavior of the tracer diffusion coefficient is different in the
five investigated geometries.
 The biggest diffusion coefficient for
most mean densities is observed in the "ladder" geometry because it
sorts particles based on their orientation while keeping them
moving. The smallest diffusion coefficient for most mean densities is observed in
the geometry with alternating pockets because it is trapping all
particles regardless of their orientation. The general conclusion from
this comparison is, that any deviation from the smooth tube is
increasing the critical density. This conclusion directly opposes
previous expectations (see for example \cite{sla_kot_25}). 

Next, we turned to analyzing densities of the separated phases. We
showed that for a given geometry and global mean density, the densities
of phases depend only on the P\'eclet number. It turned difficult to
establish the critical P\'eclet number below which no phase separation
occurs. From our data we deduce the rough upper bound
$\mathrm{Pe}_c<5$.

In the geometry with periodic pockets we observe a spontaneous ratchet
current due to the broken spatial mirror symmetry. Depending on the
average density of particles, we observe three regimes which
manifest themselves in the behavior of the current. For low enough
average density the stationary ratchet current behaves analogously to
ASEP model. Beyond a certain density, this current becomes metastable
and after some time jamming occurs, and the current drops suddenly to
a much smaller jammed current. Therefore, in this regime we can
determine two characteristic currents for a single average density,
namely the metastable and the jammed current. We found that the
jamming times obey exponential distribution, 
indicating that the transition from the metastable unjammed state to
the stable jammed state is a Poisson process. The existence of
long-lived metastable states itself means, that the transition is
driven by a nucleation mechanism and therefore is analogous to first
order equilibrium phase transition.  The mean jamming time diverges
when the average density approaches a critical value. On the other
hand, for high enough densities, the mean jamming
time decreases with the density exponentially, where the coefficient
in the exponent is
independent of the rate $\alpha$. These results are consistent with previous
findings \cite{sla_kot_25}.

Finally, beyond yet
larger density, the jamming time is so short that only jammed current
can be established. All three regimes described above constitute a
current-density diagram, which is characterized by two disjoint
branches, the first joining the low-density regime with metastable
currents, while the second corresponds to stationary jammed current
which continues into  current in the high-density regime. The
mean current in the jammed regime decreases linearly with average
density. The reason for this is that the current is proportional to
the free part  of the channel, 
not occupied by jammed clusters.

We obtained the current-density diagram also in the mean-field
approximation and we find good agreement for densities well below the
density at which the metastable current first occurs. Obviously, the
mean-field approximation cannot capture the jamming and breaks down
when strong spatial correlations accompanying the buildup of jammed
states start to occur. However, the mean-field approximation correctly
predicts that the ratchet current is a decreasing function of the
switching rate.

We also compared the simulations with solution of coarse-grained
hydrodynamic  equations, in the modified geometry in which the pockets
consist of long chain of sites. We found a good agreement in the
unjammed regime, but the jamming itself is not captured. Therefore,
the current-density diagram agrees for low enough densities but
departs sudddenly from the simulation results as soon as the
metastable states start to 
occur. Interestingly, for densities close to the maximum allowed
value, the hydrodynamic equations give again very good agreement. This
corresponds to situation, where the high-density homogeneous state is
stable to perturbations and no phase separation is possible.
In principle it should be possible to study time-dependent
hydrodynamic equations and see also the onset of jamming. We leave
this problem to future investigations.

To compare physical scales of our results with experiments, we looked
at the findings of Ref. \cite{ebb_tu_how_gol_12}, where the velocity
of platinum-coated 
Janus particles in hydrogen peroxide was measured. From this paper it
can be deduced that the P\'eclet number follows the formula
\begin{equation}
  \mathrm{Pe}=6.5*l_e^2\,C\,R
  \label{peclet-experim}
\end{equation}
where $R$ is the radius of the particle, $C$ concentration of the
hydrogen peroxide, and $l_e=6.2\cdot 10^{-11}\mathrm{m}$ a
characteristic length determined by the chemical reaction in question,
whose value is obtained by fitting the experimental data. Note that
this formula does not contain neither temperature nor viscosity
explicitly. However, there is an implicit temperature dependence hidden in
$l_e$ (the ultimate reason for that is the temperature dependence of chemical reaction rates), as explained in  \cite{ebb_tu_how_gol_12}.
For particles with radius $R=0.5\mu\mathrm{m}$ and $1$M solution,
the formula (\ref{peclet-experim})
gives $\mathrm{Pe}=7.5$, which is just slightly above the critical value where
MIPS appears. Increasing the concentration, we go deeper into the MIPS
regime.

\end{document}